# Strain stiffening universality in composite hydrogels and soft tissues

Jake Song,[1,2,#,*] Elad Deiss-Yehiely,[1,3] Serra Yesilata,[2] and Gareth H. McKinley[2,*]

[1]Department of Materials Science and Engineering, Massachusetts Institute of Technology
[2]Department of Mechanical Engineering, Massachusetts Institute of Technology
[3]Koch Institute for Integrative Cancer Research, Massachusetts Institute of Technology
[#]Department of Mechanical Engineering, Stanford University (current)

*Correspondence to jakesong@stanford.edu, gareth@mit.edu

Soft biological tissues exhibit a remarkable resilience to large mechanical loads, a property which is associated with the strain stiffening capability of the biopolymer networks that structurally support the tissues.[1,2] Yet, recent studies have shown that composite systems such as tissues and blood clots exhibit mechanical properties that deviate from those of the polymer matrix – demonstrating stiffening in compression, but softening in shear and tension.[3] The microscopic basis of this apparent paradox remains poorly understood. We show that composite hydrogels and tissues do indeed exhibit non-linear elastic stiffening in shear – which is governed by the stretching of the polymer chains in the matrix – and that it is driven by the same mechanism that drives compression stiffening. However, we show that the non-linear elastic stiffening in composite hydrogels and tissues is masked by mechanical dissipation arising from filler-polymer interactions known as the Mullins effect,[4] and we introduce a method to characterize the non-linear elasticity of the composites in isolation from this overall strain softening response through a large-amplitude oscillatory shear technique.[5] We thus present a comprehensive characterization of the non-linear elastic strain stiffening of composite hydrogels and soft tissues, and show that the strain stiffening in shear and compression are both governed by universal strain amplification factors that depend on essential properties of the composite system, such as the filler concentration $\phi$ and the filler-polymer interaction strength $\varepsilon$. These results elucidate the microscopic mechanisms governing the non-linear mechanics of tissues, provide design principles for engineering tissue-mimetic soft materials, and have broad implications for understanding cell-matrix interactions in living tissues under strain.[6-8]

We first study the compression stiffening of a model composite hydrogel system consisting of $Ca^{2+}$-crosslinked pectin and dextran particles – a structural mimic of soft biological tissues consisting of cells embedded in a biopolymer network[3] – and systematically vary the particle volume fraction $\phi$ with two different particle-polymer interaction strengths $\varepsilon$ (Fig. 1A). The particles are large (diameter of ~ 120 μm, Fig. S1), and can sediment readily before gelation occurs; we address this by adding Carbopol into the hydrogel, which prevents sedimentation through yield stress stabilization of the particles (Fig. S2). We perform quasi-static compression experiments on the composite hydrogels (Fig. 1B); our systems are completely compressible under these conditions, as revealed a near-zero Poisson's ratio $v$ during compression (Fig. S4). The experiments show that the pectin hydrogels undergo an increase in the linear shear modulus with increasing $\phi$, and also exhibit greater non-linear stiffening in $G'(\lambda)$, the storage modulus as a function of axial strain $\lambda$ (Fig. S5A). This compression stiffening effect is amplified in composite hydrogels with higher $\varepsilon$, as hydrogels with positively charged ($C$) particles exhibit greater compression stiffening compared to neutral ($N$) particles (Fig. S5B), an effect which is independent of both the filler concentration and the linear modulus of the hydrogel (Fig. S5C). We find that the normalized compression stiffening curves as a function of $\phi$ and $\varepsilon$ can be shifted onto a universal master curve (Fig. 1C), with the associated compressive strain amplification factor $a_\lambda$ systematically increasing with $\phi$ and $\varepsilon$ (Fig. 1D).

These findings can be contextualized with some of the recently proposed models for understanding compression stiffening in tissues and composite hydrogels. One such model is that of van Oosten et al.[3], which suggest that incompressible fillers can cause amplified stretching of the polymers under compression by restricting interstitial space. Another is that of Shivers et al., which suggest that compression causes the non-affine displacement of rigid fillers, which then causes stretching of the bonded polymers in a percolated filler-polymer network, and eventually jamming of the fillers themselves.[9] Our results provide experimental support to the model of Shivers et al., as we observe clearly demarcated regimes of bending, stretching, and jamming in our master curve that are predicted in their simulations (Fig. 1C).

The agreement with the computational predictions of Shivers et al., provides an appropriate context to understand the higher compressive strain amplification factor $a_\lambda$ observed in composite hydrogels with higher $\varepsilon$. The onset of the steep increase in $G'(\lambda)$ at low $\lambda$ and the reduction in slope at high $\lambda$ can be attributed to compression-induced structural transitions of the filler network – percolation and jamming, respectively. While the percolation threshold is difficult to isolate in the system as the polymer network alone also exhibits stiffening at large $\lambda$, the jamming threshold can be isolated from the curve as occurring approximately at $\lambda_{jam} = 1.25/a_\lambda$ (Fig. 1C). A quantitative evaluation of this inflection point over two sets of experiments allows us to characterize the particle jamming threshold $\phi_{jam} = \phi/(1 - \lambda_{jam})$, yielding $\phi_{jam}(C) = 0.49$ and $\phi_{jam}(N) = 0.82$ (Fig. S7). Given the size and stiffness invariance in the $C$ and $N$ particles, we interpret this difference in $\phi_{jam}$ as arising from the larger effective volume fraction $\phi_{eff}$ (and effective radius $r_{eff}$) of the more attractive $C$ particles relative to the $N$ particles. The $r_{eff}$ of the $C$ particles relative to the $N$ particles, $\chi$, can be calculated via the relation $\chi = r_{eff}(C)/r_{eff}(N) = \left(\phi_{jam}(N)/\phi_{jam}(C)\right)^{1/3} = 1.18$.

These findings are corroborated by comparing the linear viscoelastic shear modulus of the hydrogels at zero axial strain $G'(0)$, which are systematically higher for hydrogels with $C$ fillers than $N$ fillers at all $\phi$. We find the filler-induced reinforcement effect in our hydrogels to be substantially greater than that expected from the phantom network theory.[10] We thus fit the results to a more classical reinforcement model of the form:

$$\lim_{\gamma \to 0} \frac{G'(0)}{G'_{pec}(0)} = 1 + 2.5\phi_{eff} + c_2 \phi_{eff} \quad (1)$$

where $c_2$ is the second virial coefficient which can vary from 2.5 to 15.6,[11] and $\phi_{eff} = (r_{eff}/r)^3 \phi$. This model is often referred to as the hydrodynamic model as it is analogous to the expression used to describe the hydrodynamics of particle suspensions.[12] The relation cannot be used to quantify the absolute volume fractions in our systems as the relation neglects contributions of multi-body particle interactions at high $\phi$, and as different values of $c_2$ can influence $\phi_{eff}$. Regardless, the model still provides insight into the relative difference in $r_{eff}$ of the $C$ and $N$ filled systems, which we find to be $\chi = 1.16$ using $c_2 = 14.1$ by Guth and Gold (Fig. 1E),[13] and $\chi = 1.17$ using $c_2 = 5.2$ by Batchelor and Green (Fig. S8),[14] both of which are in excellent agreement with $\chi = 1.18$ obtained from the jamming threshold measurements.

The increase in $\phi_{eff}$ with increasing $\varepsilon$ in our systems can be attributed to the greater polymer adsorption capacity of the attractive $C$ particles compared to the $N$ particles (Fig. S9). This increased polymer adsorption can cause the formation of a dense layer of immobile polymers on the particle surface, which effectively increases the size of the fillers[15-17]. Moreover, increased polymer adsorption can increase bridging interactions between particles, which results in the densification of the bridged particle network (via a decrease in the interparticle distance).[18] Indeed, we can observe such bulk contractions in the $C$ filled gels compared to the $N$ filled gels from normal force measurements during gelation (Fig. S10). Importantly, we find that the compressive strain amplification factor $a_\lambda$ shows a good collapse when plotted as a function of $\chi^3 \phi$ (Fig. 1D), where $\chi$ is the effective radius of the $C$ fillers relative to the $N$ fillers (i.e., $\chi(C) = 1.16$, $\chi(N) = 1$). This shows that a difference in $\phi_{eff}$ can adequately explain the strain amplification observed in composite hydrogels with different filler-polymer interaction strength $\varepsilon$.

Composite hydrogels and soft tissues have been shown to exhibit strain softening in tension and shear, in contrast to the stiffening observed in compression shown thus far.[3,19] This behavior appears paradoxical given the dramatic tension and shear stiffening exhibited by their constituent biopolymer matrix.[1,2] This tension and shear softening behavior has rationalized by van Oosten et al.[3] through a mechanism in which incompressible fillers shield the polymer matrix from strain. However, this form of tension and shear strain softening effect has been observed more generally in polymeric composites for many decades, and is canonically described as the Payne effect[20] (in which the storage modulus $G'$ decreases with increasing strain amplitude $\gamma_0$, and in which the loss modulus $G''$ exhibits an overshoot at the yield point) and the Mullins effect[4] (in which stress softening occurs with increasing cyclic strain), depending on the experimental protocol. We tested the non-linear shear properties of our composite hydrogel via large-amplitude oscillatory shear (LAOS), and indeed observed a strain softening response which is consistent with these past observations. While pectin hydrogels without fillers exhibit strain stiffening (as evidenced by an increase in $G'$ with increasing $\gamma_0$) pectin hydrogels filled with $C$ and $N$ particles do not exhibit any form of stiffening in shear (Fig. 2A). However, since the raw stress waveform becomes distorted in LAOS experiments, the $G'$ and $G''$ computed by the rheometer (which considers just the first-harmonic response of the stress and strain, and are denoted $G'_1$ and $G''_1$) provide an oversimplified description of the non-linear mechanics of both systems (Fig. 2B,C).[21] An alternative method of analysis is clearly needed to extract non-linear elasticity in these systems.

One accepted method is to compute the differential storage modulus at the maximum strain of each cycle, $G'_K(\gamma_0) = (d\sigma'/d\gamma)|_{\gamma_0}$, where $\sigma'$ is the elastic component of the total stress (Fig. 2B).[22] This quantity is the LAOS-equivalent of the differential storage modulus $K'(\sigma_0)$ obtained by superimposing a small-amplitude oscillatory shear on a step stress $\sigma_0$, a protocol frequently used to characterize the non-linear elasticity of biopolymer networks.[23,24] Analysis of $G'_K(\gamma_0)$ for our systems reveals shear strain stiffening in the composite hydrogels, but to a lower extent than that exhibited by the filler-free pectin hydrogels (Fig. S11A). The raw stress-strain Lissajous curves of the pectin hydrogels and composite hydrogels provide clues on why composite hydrogels exhibit lower $G'_K(\gamma_0)$ values than pectin hydrogels: while the $\sigma'(t)$ curves of the pectin hydrogels are superposed on each other at all $\gamma_0$ until reaching the yield strain (the point of significant decrease in $G'_1$ with $\gamma_0$, Fig. 2A), the $\sigma'(t)$ curves of the composite hydrogels are horizontally translated at input strains below the yield strain (Fig. 2B,C). These results show that inelastic dissipation occurs prominently in the composite hydrogels prior to yielding, thus obfuscating measurements of non-linear elasticity in these systems.

These inelastic effects can be attributed to plastic deformations that occur in the gels prior to macroscopic yielding. Evidence for these effects can be seen in the overshoot in the $G''$ in the composite hydrogels prior to yielding, which is a hallmark property of the Payne effect[20] and a measure of the irrecoverable plastic deformation that occurs in the microstructure in the non-linear regime.[25] Further evidence for these effects are revealed by cyclic compression and transient LAOS experiments on the composite hydrogels, which show that subsequent loading curves memorize and trace the previous unloading curves, with increasing irreversible plastic deformations per loading cycle (Fig. S12) – a hallmark property of the Mullins effect.[4] The Mullins effect therefore leads to a horizontal translation of $\sigma'(t)$ – as seen also in and a softening effect that is convolved in measurements of $G'_1$ (Fig. 2A) and $G'_K(\gamma_0)$ (Fig. S12). In the absence of such plastic effects, we may expect the non-linear elasticity of the pectin hydrogels and composite hydrogels to be coincident. This would correspond to the physical scenario predicted by Shivers et al., in which plastic effects are absent (as polymers are permanently bound to the particles), and in which tension causes stiffening via stretching of the bridging polymers.[9]

We show that the Mullins effect observed in LAOS (Fig. 2C, Fig. S12) can be eliminated to obtain the true non-linear elastic response of the system by computing the *strain-dependent differential storage modulus*, $G'_K(\gamma) = d\sigma'/d\gamma$ (Fig. 2C inset). When represented as a function of the input oscillation amplitude $\gamma_0$, the $G'_K(\gamma)$ curves are self-similar for both pectin and composite hydrogels. However, the latter results become horizontally off-set with increasing $\gamma_0$ due to the Mullins effect (Fig. 2D,E). The translated $G'_K(\gamma)$ curves can thus be horizontally shifted onto a reference $G'_K(\gamma)$ curve in the linear regime to construct a master curve for each system,[26,27] via a shift factor which characterizes the strain-dependent plasticity in our system due to the Mullins effect, which we refer to as the Mullins factor $M_f$. The $M_f$ values are near unity for the pectin hydrogels at all $\gamma_0$ (minimal plastic effects), but they increase in composite hydrogels with increasing $\phi$ and $\varepsilon$ (Fig. 2F). The observation of greater $M_f$ in attractive hydrogels with $C$ particles compared to $N$ particles is also supported by the differences in the Mullins effect observed in cyclic compression experiments for these two systems (Fig. S12B-E). Overall, these results suggest that stronger filler-polymer interaction strength $\varepsilon$ result in greater plastic dissipation in composite hydrogels. This can be rationalized by the greater degree of immobilization of the polymers on the particle surface with increasing $\varepsilon$ (Fig. 1F),[16,17] as the Mullins and Payne effects can be attributed to the strain-induced detachment of the immobilized polymers on the particle surface which bridge the particles,[28,29] It is noted that the shift factor data for composite hydrogels with $N$ particles are truncated at higher $\gamma_0$ due to the system exhibiting clear signatures of slip before yielding occurs. This is identified by a gradual drop in $G_1'$ and $G_1''$ prior to yielding (Fig. 2A, S14A)[30] and a more softened response in the $G_K'(\gamma)$ curve relative to those in systems without slip (Fig. S14B,G). We find neither the pectin hydrogels nor the composite hydrogels with $C$ fillers affected by these slip-related artifacts during measurement.

The shifted $G_K'(\gamma)$ curves provide an unadulterated characterization of the non-linear elasticity of composite systems (Fig. S15A). The shifted $G_K'(\gamma)$ curves for pectin and for composite hydrogels as a function of $\phi$ and $\varepsilon$ are self-similar, and can be shifted onto a single master curve using a shear strain amplification factor $a_\gamma$ – in similar vein as the results from compression experiments (Fig. S15B). In fact, we find that both the compression and shear stiffening of composite hydrogels as a function of $\phi$ and $\varepsilon$ can be described by a universal master curve (Fig. 2G), with coincident values of $a_\lambda$ and $a_\gamma$ (Fig. 2H). These results demonstrate the universal non-linear elastic response of composite hydrogels in compression and shear, which is governed by the stretching of the matrix polymer chains. This observation is consistent with simulation results of Shivers et al., who observe a similar universality in compression and tension in composite systems without plastic effects.[9] We furthermore find that the strain amplification factors $a_\lambda$ and $a_\gamma$ can be reasonably described in a parameter-free manner with a simple 1-D filler-induced affine strain amplification model (Fig. 2I)[31]:

$$a_i / a_i(0) = 1 / (1 - \phi_{eff}^{1/3}) \qquad (2)$$

where $i = \lambda$ or $\gamma$. This model suggests that fillers effectively enhance the strain experienced by the bridging polymer matrix,[9,32] an effect which is amplified with increasing filler-polymer interaction strengths due to the increase in the effective filler volume $\phi_{eff}$ in the system (Fig. 1F).[15] Overall, these results demonstrate the critical roles played by filler volume and filler-polymer interactions – which are fundamental parameters in biological tissues and technological composites alike – in amplifying the non-linear mechanical responses of the system.

It is interesting to compare the extent of strain amplification with the linear viscoelastic shear modulus with increasing filler loading (Fig. 1E). We find that $a_\lambda$ and $a_\gamma$ trace the linear shear modulus $G'/G'_{pec}$ at $\phi \leq 0.2$, suggesting that strain amplification can provide an empirical description of the increased elasticity of composite systems at low filler concentrations (Fig. S17), even if it does not account for some of the pertinent physics arising from filler incorporation such as the hydrodynamic reinforcement effect.[11,33,34] However, beyond $\phi > 0.2$ we find that $G'/G'_{pec}$ exhibits a steep increase and deviates from $a_\lambda$ and $a_\gamma$ (Fig. S17). This steep increase in $G'/G'_{pec}$ can be explained by the emergent elasticity of a percolated particle-polymer network,[17,29,35,36] the presence of which is evidenced in the hallmark behaviors of our composite hydrogels such as compression stiffening,[9] and plastic dissipation at large $\gamma_0$.[28]

Our experimental results and the simulation results of Shivers et al.[9] demonstrate a strain stiffening universality in composite hydrogels where stiffening is equally amplified in shear and in compression by essential parameters such as $\phi$ and $\varepsilon$. However, both our work and those of Shivers et al. demonstrate this effect on model composites which are structurally different to tissues. The model of Shivers et al. utilize fiber networks with rigid inclusions, while our system utilizes polysaccharides also with relatively stiff inclusions (~ $10^6$ Pa).[37] These model systems are not equivalent to soft tissues which are multi-polymer composites of flexible proteoglycans and fibrous polymers embedded with *cells*, which have a low stiffness ranging from $10^2 \sim 10^5$ Pa,[38] and which exhibit matrix-stiffening contractile activity.[39] Thus, the broad implications of strain stiffening universality remains unclear in soft biological tissues, as mostly softening responses have been observed in tissues under tension and shear thus far[3,19].

To explore this, we perform compression stiffening experiments on mouse tissues: lung, heart, adipose, liver, and partially-decellularized (PD) liver[40] (Fig. 3A, S16). We find the tissue samples to be more incompressible than the composite hydrogels, with $\nu$ ranging from 0.2 to 0.4, which is indicative of their low poroelasticity relative to the pectin hydrogels (Fig. S19).[41] All measured $G'(\lambda)$ data can be shifted onto the lung data which shows the weakest compression stiffening (Fig. 3B). The first harmonic storage modulus $G'_1$ collected from LAOS experiments on the tissues show an even more pronounced Payne effect (Fig. 3C) compared to the composite hydrogels (Fig. 2A), and correspondingly, we observe significant dissipation occurring at large strains as indicated by the horizontal translations in the Lissajous curves (Fig. 3D). The Lissajous curves also show significant strain-stiffening behavior in $\sigma'(t)$, which we can now extract via our $G'_K(\gamma)$ protocol.

To our knowledge, this represents the first isolated characterization of the non-linear elasticity of biological tissues. As such, we first analyze $G'_K(\gamma)$ as a function of elastic stress $\sigma'$, from which we can gain molecular insights into the origin of non-linear elasticity in tissues. There is extensive literature on the origins of the scaling of the differential modulus with stress, i.e. $G'_K(\gamma) \sim \sigma'^z$. The most common interpretation is that this scaling arises due to the force-displacement response of the individual polymer strands.[2,23,42] In this case, $z$ follows the force-extension (F-L) relationship of individual polymer or filament strands (which can be characterized by atomic force microscopy experiments), with $dF/dL \sim F^z$.[23] The most common value of $z$ in biopolymer systems is the case when the constituents obey entropic elasticity predicted by the worm-like chain (WLC) model,[43] in which case $z = 1.5$. However, this number can be lower in biopolymer systems such as polysaccharides which are more extensible or have enthalpic components to the elasticity.[44] One such example is pectin, and an analysis of the force-extension relationship of pectin strands[45] reveal a scaling of $dF/dL \sim F^{1.1}$ (Fig. S20A), which is in excellent agreement with the scaling of $G'_K(\gamma) \sim \sigma'^{1.1}$ in our composite systems (Fig. S20B). This shows that polymer chain stretching underlies the non-linear elasticity of pectin hydrogel and the composites. This analysis also shows that a similar mechanism might be responsible for the $z < 1.5$ scaling of the differential modulus observed in other polysaccharide gels in literature, for example agarose[46] and chitosan.[47]

In tissues, we may expect tissues to exhibit a scaling of $G'_K(\gamma) \sim \sigma'^{1.5}$ because most of the biopolymer constituents including actin,[48] collagen molecules,[49] and proteoglycans[50] follow the WLC model. On the other hand, tissue mechanics are often assumed to be dominated by a fibrillar collagen network in the connective tissue extracellular matrix (ECM) scaffold. Collagen fibers are hierarchical assemblies of collagen molecules, have persistence lengths on the order of cm,[51] and do not follow WLC behavior.[52] Reconstituted collagen fiber networks show a scaling of $z < 1.5$ in the differential modulus, which is often attributed to the stress-induced stabilization of the subisostatic fiber network.[53,54]

We find that all our tissues – both whole organs and the connective tissue adipose – follow a scaling of $G'_K(\gamma) \sim \sigma'^{1.5}$ (Fig. 3E). This suggests that non-linear shear strain induces stretching of polymer chains and filaments in the tissues, rather than directly engaging the non-linear response of the collagen fiber network in the ECM. Prior studies of tension experiments on rat tendon have shown that the total strain of the whole tendon is greater than those experienced by the collagen fibers inside the tendon, thus suggesting that the strain may primarily occurs in the proteoglycan-rich matrix.[52,55] Our results reinforce this observation across different whole organs and connective tissues. The observation of $G'_K(\gamma) \sim \sigma'^{1.5}$ also allows us to rule out the contribution of slip on the non-linear elasticity of tissues, as $z = 1.5$ corresponds to the upper limit of the theoretical expectations, and sample slip has been shown to cause a reduction in $z$ (Fig. S14B,G).

We also find signatures of substantial plastic dissipation in the characterized tissues, evidenced by the Payne effect (Fig. 3C) and the horizontal translation of the Lissajous curves with $\gamma_0$ (Fig. 3D). However, the resulting Mullins factors $M_f$ of the tissue samples are noticeably higher than that observed in the composite hydrogels (Fig. 3F, S23B), with no apparent correlation with the onset strain for strain stiffening as was previously observed in composite hydrogels (as evidenced by the lung exhibiting the greatest $M_f$). This indicates the presence of dissipative mechanisms in tissues other than those governing our composite hydrogel systems (Mullins effect). The whole tissues studied here are architecturally complex, and could exhibit other dissipative mechanisms such as scissions of covalent bonds in the ECM, or detachment of cell-cell junctions in epithelial layers of the tissues. We also cannot rule out dissipative contributions from slip as the tissue samples are loaded *ex situ*; however, we note that we do not find signatures of slip from the observed non-linear elasticity $G'_K(\gamma)$ of the tissues, and that similar extents of horizontal translation in the Lissajous curves at large strains have been reported in tissue samples rigidly bound with cyanoacrylate glues.[56,57]

We now compare the shear measurements of tissues with compression measurements. We find that the onset strain for shear stiffening are in agreement with the onset strain for compression stiffening (Fig. S22A). We also find that the $G'_K(\gamma)$ data can be shifted onto a master curve with the shear strain amplification factor $a_\gamma$ (Fig. S22B), in similar vein as the compression data (Fig. 3B). Thus, similar to that observed in composite hydrogels (Fig. 2G), we observe a universal master curve for various tissues as a function of compression and shear (Fig. 3G). Furthermore, we observe that the strain amplification factors $a_\lambda$ and $a_\gamma$ (shifted with reference to the lung data) are practically equivalent for all tissues, and that they also deviate from $G'(0)/G'_{lung}(0)$ (Fig. 3H) – both findings in agreement with observations in our composite hydrogels.

In total, these results demonstrate the strain stiffening universality in tissues, which shows that a filler-induced strain amplification mechanism controls the non-linear elasticity of tissues in shear and in compression. The universality of this effect in both tissues (Fig. 3G) and composite hydrogels (Fig. 2G) confirms the idea that tissue mechanics can be modeled by a composite hydrogel consisting of strain-stiffening polymers and attractive fillers.[3,9] We have demonstrated the importance of key composite variables such as filler volume $\phi$ and filler-polymer interaction strength $\varepsilon$ in mediating the mechanical response of composite hydrogels, and these findings in turn suggest that variables such as cell density and cell-ECM interactions may play key roles in dictating the non-linear mechanics of living tissues. Living tissues also exhibit additional complexities not found in composite hydrogels that could contribute to their non-linear mechanics, such as directional densification and alignment of the ECM due to contractile activity[39,58-60], softness[38] and poroelasticity[61] of the cells, interpenetrating biopolymers in the ECM,[62] and cell-cell adhesions and aggregates as found in epithelial tissues and tumors. Deciphering the roles played by these different mechanisms on the non-linear response of tissues could provide important insights into cell-ECM interactions that mediate development[7] and disease,[8,63] under mechanical forces, and offer guidance on designing tissue-mimetic soft materials.

**Acknowledgement**

J.S. acknowledges financial support from the MIT Lemelson-Vest award and the MIT MathWorks fellowship. J.S. and G.H.M acknowledge helpful discussions with P. Janmey (U Penn), I. Dellatolas (MIT), I. Bischofberger (MIT), E. Del Gado (Georgetown), and J. Shivers (U Chicago).

**Methods**

**Composite hydrogel.** The hydrogels were prepared using similar protocol previously reported for alginate[64], by mixing aqueous solutions of low-methoxy pectin (Unipectin 700, Cargill) with $CaCO_3$ at a stoichiometric ratio of 1 $Ca^{2+}$ : 2 $COO^-$ present in the galacturonic acid groups. Carbopol (940, Lubrizol) and dextran particles (Sephadex A-25 and G-25, Cytiva) were then added to the system, and lastly gluocono-D-lactone (GDL) was added at a stoichiometric ratio of 1 $Ca^{2+}$ : 3 GDL to induce $Ca^{2+}$ dissolution and gelation. The composite hydrogels had a final composition of 1 wt. % pectin, 0.5 wt. % Carbopol (see flow curve in Fig. S2), and varying wt. % of dextran to meet a specified target volume fraction of their swollen state $\phi$ (see details in Fig. S1).

**Tissues.** All animal experiments were approved by the Massachusetts Institute of Technology Committee on Animal Care, protocol 0720-048-23. Tissues were collected from 20 week old outbred CD-1 mice (Charles River Laboratory) that were euthanized by $CO_2$ asphyxiation and confirmed via a cervical dislocation. Lung, heart, liver, and omentum adipose tissue were collected from the mice, stored in PBS solution on top of ice, and analyzed within six hours of euthanasia and collection (sample pictures in Fig. S18). Partially decelluarized (PD) liver were prepared by immersing a liver sample in 2 % Triton-X solution for 24 hours to wash the cell membrane from the tissue, following adaptations of literature protocol[3].

**Rheology.** Compression and shear experiments were conducted using the Anton Paar MCR-302 and the TA DHR-3 rheometer. Compression experiments were done by performing a $\lambda = 0.025$ compressive step strain every 500 s, superposed by a small-amplitude oscillatory shear (SAOS) with an imposed shear strain of $\gamma_0 = 0.005$. Large-amplitude oscillatory shear (LAOS) experiments were performed by collecting the steady-state waveforms of the stress-strain response at each $\gamma_0$. The elastic stress $\sigma'(t)$ and viscous stress $\sigma''(t)$ contributions to the total stress $\sigma(t)$ in an oscillatory strain waveform $\gamma(t) = \gamma_0 \sin(\omega t)$ were isolated from the total stress response using MITlaos.

The mixtures to form composite hydrogels were loaded onto the MCR-302 (25 mm parallel plate) or the DHR-3 (20 mm parallel plate), and allowed to undergo gelation for 3 hours before compression or shear characterization (gelation profile in Fig. S24). Regular stainless steel plates were used for characterizing composite hydrogels as we observed minimal differences when using sandpaper (Fig. S13). Tissue samples were loaded onto the MCR-302 (10 mm parallel plate) or the DHR-3 (8 mm parallel plate) which were coated with sandpaper (Trizact A10, 3M) to minimize slip effects. Tissue samples were loaded until a normal force of 0.1 N was reached, and trimmed around the rim of the plate before proceeding with analysis. The outermost rim of the composite hydrogel and tissue samples were then sealed with mineral oil to prevent dehydration during testing.

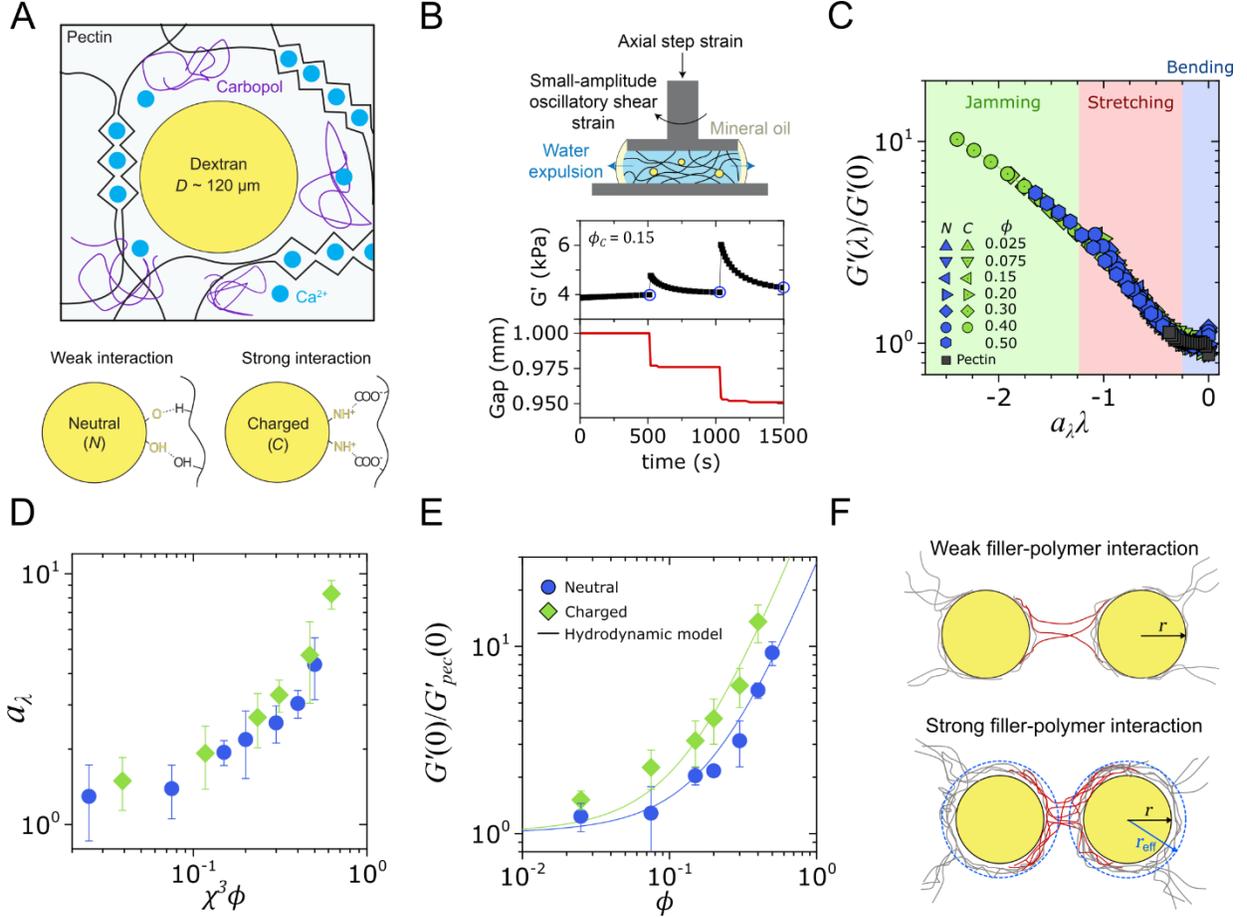

**Figure 1. Compression stiffening universality in composite hydrogels.** A) Illustration of the composite hydrogel microstructure. The system consists of low-methoxy pectin cross-linked by $Ca^{2+}$ ions, Carbopol, and dextran particles that undergo weak (H-bonding) and strong (ionic) interactions with the anionic pectin chains. The diameter $D$ of the particles is measured via imaging (Fig. S1). B) Illustration of the experimental protocol, adapted from reference[3]. Time-dependent small-amplitude oscillatory shear (SAOS) is superposed on a step axial strain $\lambda$. The resulting storage modulus $G'$ is monitored for 500 s (to allow poroelastic equilibration[65]), and the last data points at each step (blue circle) are reported. The hydrogels are fully compressible under these conditions, as shown in Fig. S3. C) Representative master curve of normalized $G'$ (relative to the modulus at zero compressive strain, $G'(0)$) as a function of $\lambda$ rescaled by the compressive strain amplification factor $a_\lambda$, for the composite hydrogel system as a function of $\phi$ and $\varepsilon$ (see Fig. S6 for repeat set of data). The corresponding regimes of bending (softening in $G'$ with $\lambda$), stretching (sharp increase in $G'$ with $\lambda$), and jamming (decrease in slope of $G'$ with $\lambda$) illustrated by Shivers et al.,[9] are highlighted. D) Compressive strain amplification factor $a_\lambda$ as a function of $\chi^3\phi$, where $\chi(C) = r_{eff}(C)/r_{eff}(N) = 1.16$, and $\chi(N) = 1$. Statistical features are obtained from $a_\lambda$ collected across two sets of experiments, thus providing us with $n = 4$ (Fig. 1C and S6). E) Linear elasticity of the composite hydrogels at no axial strain $G'(0)$, normalized to that of the pectin hydrogel $G'_{pec}(0)$, and fitted to the Guth-Gold ($c_2 = 14.1$) hydrodynamic model in Eqn. 1, from which $r_{eff}$ (and thus $\chi$) is obtained ($n = 4$). F) Schematic illustration of the fillers with weak ($N$) and strong ($C$) polymer interactions. Strong filler-polymer interactions lead to a thicker layer of adsorbed polymers,[15] smaller inter-particle distances (due to stronger particle-particle bridging interactions)[18], and thus a larger effective volume $\phi_{eff}$ of the fillers in the hydrogel network.

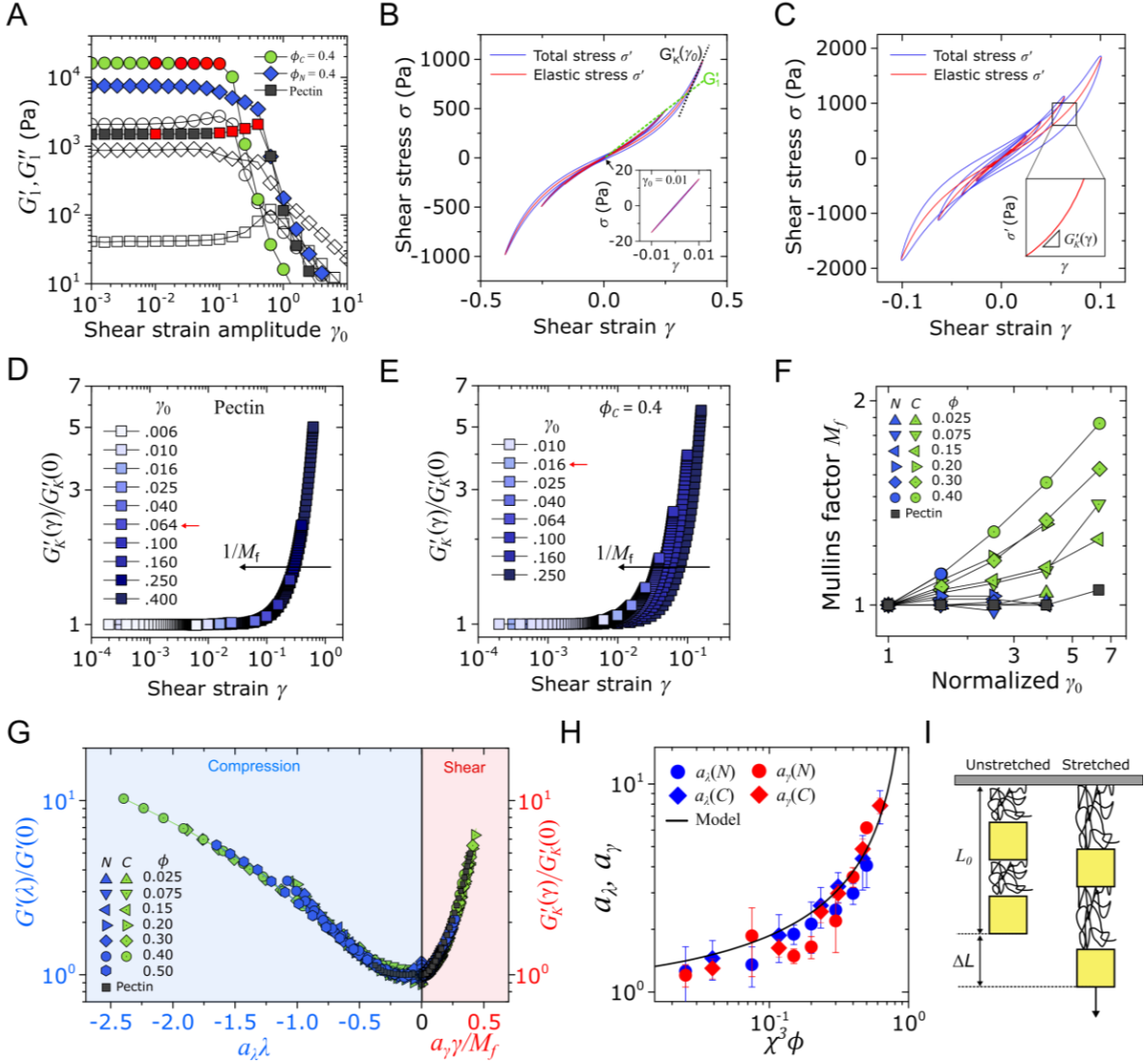

**Figure 2. Shear stiffening universality in composite hydrogels.** A) The first-harmonic storage modulus $G'_1$ and loss modulus $G''_1$ as a function of oscillatory shear strain amplitude $\gamma_0$ for pectin and representative composite hydrogels. B), C) Representative Lissajous curves of pectin and $\phi_C = 0.4$ hydrogels at $\gamma_0$ corresponding to those highlighted in red symbols in panel A. The Lissajous curves are ellipsoidal in strain at low $\gamma$ (panel B inset), but become distorted in the non-linear regime. The first-harmonic storage modulus $G'_1$ is computed from a linear fit from the origin (green dashed line), while the differential storage modulus $G'_K(\gamma_0)$ is computed from the tangent slope at $\gamma_0$ (black dotted line), both illustrated in panel B. Our protocol for calculating the *strain-dependent differential storage modulus* $G'_K(\gamma)$ is highlighted in panel C. D), E) The $G'_K(\gamma)$ of the pectin and $\phi_c = 0.4$ gels, normalized by $G'_K(\gamma = 0)$. All data are subsequently shifted to a reference curve measured at $\gamma_0$ at which the $G'_K(\gamma)/G'_K(0)$ exhibits a 5% increase (red arrow), and show an exact overlay (Fig. S15). F) The Mullins factor $M_f$ for the pectin and composite hydrogels as a function of $\gamma_0$ (normalized to the reference $\gamma_0$) for each system. G) Universal master curve for the composite hydrogels in compression and in shear, and H) corresponding strain amplification factors $a_\lambda$ and $a_\gamma$ as a function of $\chi^3\phi$, where $\chi(C) = r_{eff}(C)/r_{eff}(N) = 1.16$, and $\chi(N) = 1$. Statistics for the shear experiments ($n = 4$) are obtained from two sets of experiments, (master curve and Mullins factors of repeat experiments in Fig. S16). I) Illustration of the 1-D affine strain amplification model[31] highlighted in panel H. Total strain is defined as $\lambda_{tot} = \Delta L/L_0$, but the polymeric components of the system experience a strain of $\lambda_{pol} = \Delta L/L_0(1-\phi^{1/3})$. Thus, strain is amplified by a factor $a = \lambda_{pol}/\lambda_{tot} = 1/(1-\phi^{1/3})$ (Eqn. 2).

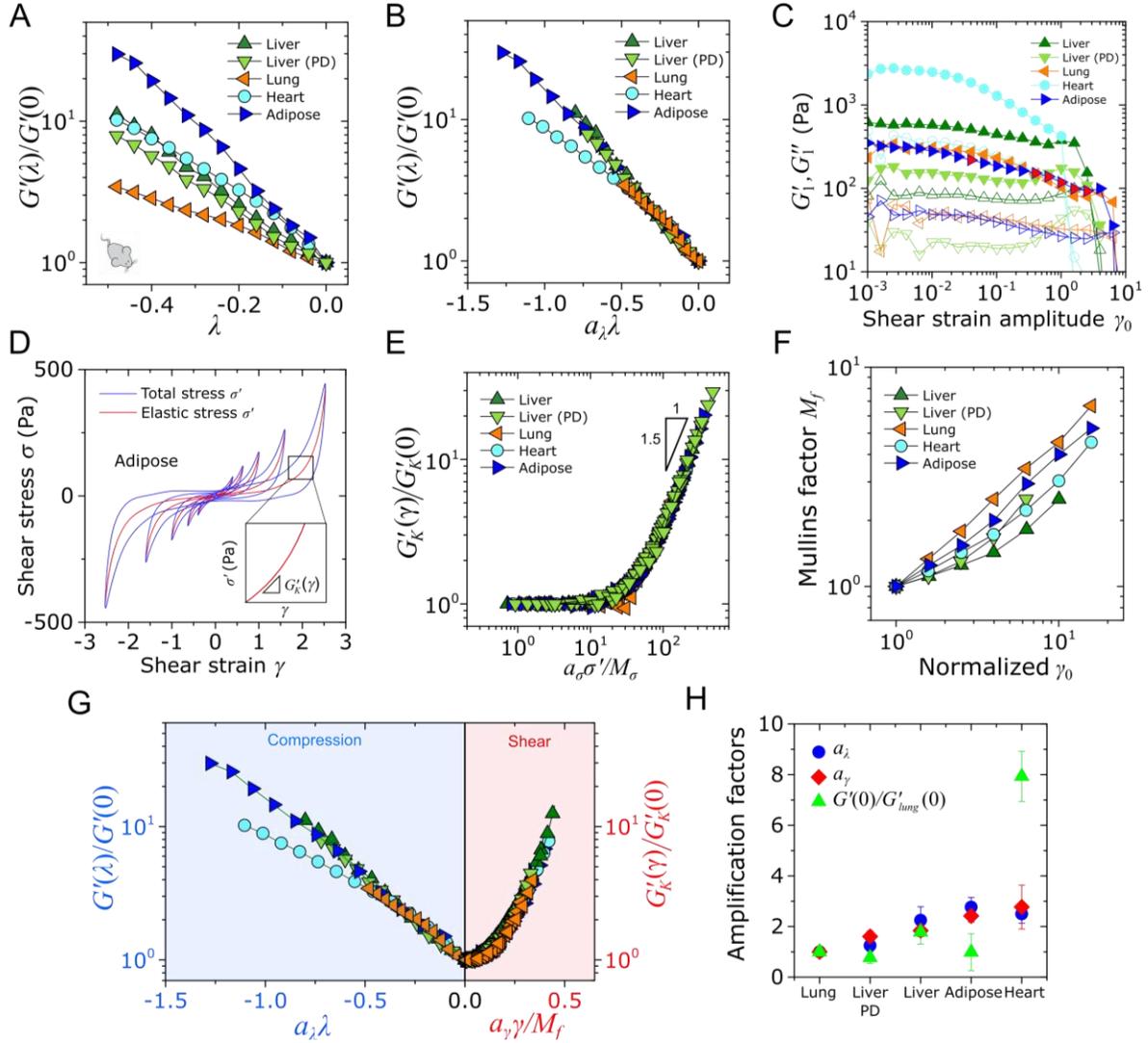

**Figure 3. Compression and shear stiffening universality in soft biological tissues.** A) Compression stiffening of soft tissue samples from mice (representative pictures in Fig. S18), which is B) shifted to the least stiffening lung sample using the compressive strain amplification factor $a_\lambda$. The heart data shows an inflection in the curve with $\lambda$, which can be attributed to the round shape of the sample (Fig. S18A) that causes bulging resistance at initial stages of compression (Fig. S19D). C) The first-harmonic storage modulus $G'_1$ and loss modulus $G''_1$ as a function of oscillatory shear strain amplitude $\gamma_0$ for tissue samples. D) Representative Lissajous curves of adipose tissue at $\gamma_0$ corresponding to the red symbols in panel C. Inset illustrates the calculation protocol for the strain-dependent differential storage modulus $G'_K(\gamma)$. E) The master curve of $G'_K(\gamma)$ for different tissues as a function of the elastic stress $\sigma'$ (see Fig. S21 for discussion of the related shift factors $a_\sigma$ and $M_\sigma$). F) Mullins factor $M_f$ for the different tissue samples (see Fig. S23B for repeat data). G) Universal master curve for the tissue samples in compression and in shear, and H) the corresponding strain amplification factors $a_\lambda$ and $a_\gamma$, plotted alongside $G'(0)/G'_{lung}(0)$ ($n = 4$). All curves are shifted to the lung reference curve. Statistical features are obtained from shift factors collected across two sets of experiments, thus providing us with $n = 4$.

# SUPPLEMENTARY INFORMATION

## Table of Contents

## Determination of added dextran mass to the composite hydrogels

The requisite mass of dextran particles to be added were determined based on the required volume fraction $\phi$ of the swollen particles. The swelling ratio of dextran was determined using two independent methods.

The first method involved calculating the mass difference between dry and swollen particles. To weigh the swollen particles, dextran was allowed to swell in $H_2O$ overnight, after which the solution was centrifuged (5000 g, 10 mins) and poured on a wet filter paper (Whatman Grade 1) on which the mass balance has been zeroed. Excess water was allowed to percolate, and then the final mass of the swollen dextran was determined ($n = 3$). The mean density value of dextran reported in literature ($\rho = 1.09$ gcm$^{-3}$),[1] to determine the final volumetric swelling ratio of dextran (443 %).

The second method involved calculating the volume ratio between dry and swollen particles by microscopy imaging. The swollen particles were imaged after swelling dry particles in $H_2O$ for 3 hours (same duration as gelation time). Calculation of the swelling ratio based on the differences in the particle sizes yielded an average swelling ratio of 422 %, in reasonable agreement with the mass method. Composite hydrogels with $\phi = 0.15$ were also imaged to determine the particle swelling ratio in the pectin gel; this control experiment reveals that the swelling ratio of the particles is negligibly affected by the particles being inside a hydrogel (Fig. S1).

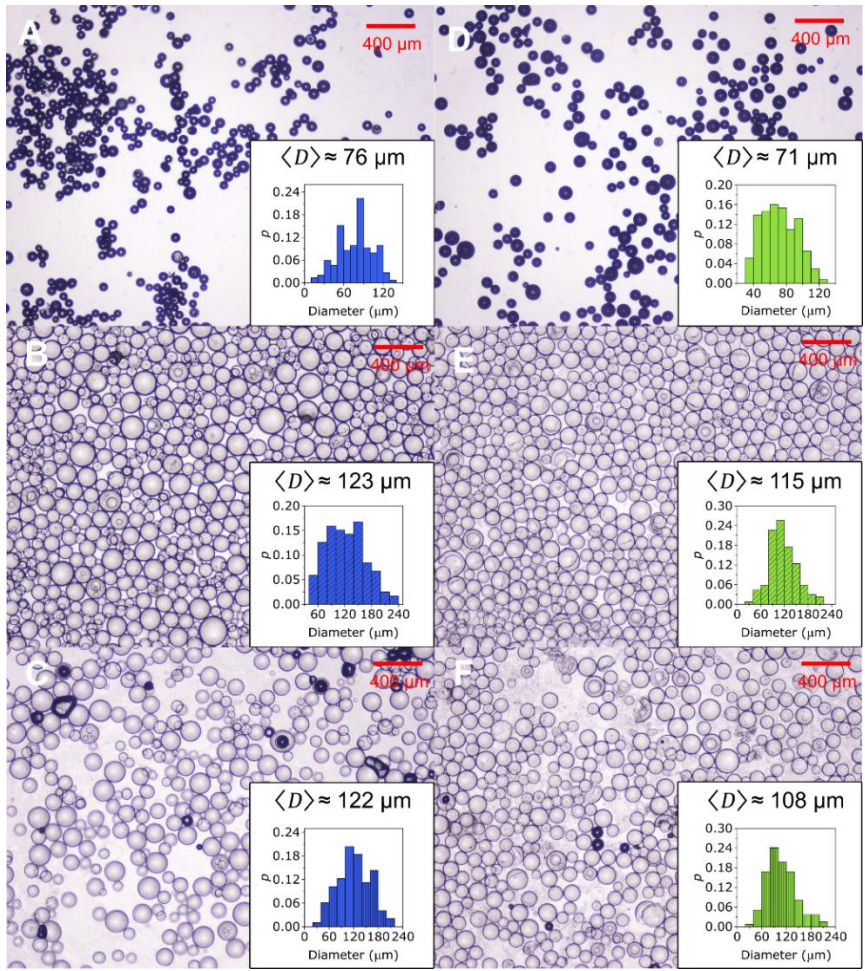

**Figure S1.** Representative images (Olympus CKX53) of neutral dextran particles in the A) dry and B) swollen state, and C) in a $\phi = 0.15$ gel state, and cationic dextran particles in the D) dry and E) swollen state, and F) in a $\phi = 0.15$ gel state. Note that the $\phi$ values refer to the concentration of dextran added to make the gel, and not the $\phi$ in the image (as the gels were imaged by compressing the hydrogels between microscope slides). Representative particle size distributions ($n = 100$) as well as the associated mean diameter $\langle D \rangle$ are shown in the inset.

## Stabilizing dextran particles from sedimentation in the composite hydrogels

We found that the addition of a yield stress fluid to be essential in creating hydrogels with stabilized particles. In the absence of yield stress fluids, we found that the particles exhibited sedimentation during gelation (Fig. S2A). Particle sedimentation can significantly affect compression stiffening in the composite hydrogels, as compressive strain will mainly result in a particle jamming response. To prevent the sedimentation of the dextran particles, we incorporated Carbopol 940 (0.5 wt. % total concentration) in the hydrogel, which results in the stabilization of the particles during gelation (Fig. S2B).

The Carbopol solution prevents dextran particle sedimentation by resisting the gravitational force of the particles. The yield stress of Carbopol at 0.5 wt. % can be analyzed through a flow curve analysis (Fig. S2C). A fit to the Herschel-Bulkley model:

$$\sigma = \sigma_y + K\dot{\gamma}^n \qquad (0.1)$$

where $\sigma_y$ is the yield stress, $K$ is the consistency index, and $n$ is the flow index, shows that the Carbopol solution has a yield stress of $\sigma_y = 0.93$ Pa. The gravitational force of the dextran particles $F$ can be computed by the relation $F = (4/3)\pi g r^3 (\rho_{dex} - \rho_{H_2O})$, where $g = 9.8$ ms$^{-2}$ is gravitational acceleration, $r \approx 60$ μm is the radius of the dextran particles (see Fig. S1), and $\rho_{dex} = 1090$ kgm$^{-3}$ is the density of dextran, and $\rho_{H_2O}$ is the density of water. $F$ can now be compared to the yield stress of the surrounding fluid, $\sigma_y$, via the relation:

$$Y_g = \frac{2\sigma_y \pi r^2}{F} \qquad (0.2)$$

where $Y_g$ is the gravitational yield number. The $Y_g$ of dextran particles in the Carbopol solution $Y_g = 26.3$, which is substantially higher than the Beris criterion[2], which states that spherical particle sedimentation occurs when $Y_g < 0.143$. This calculation assumes that the particles do not directly aggregate, which we discuss in the next section.

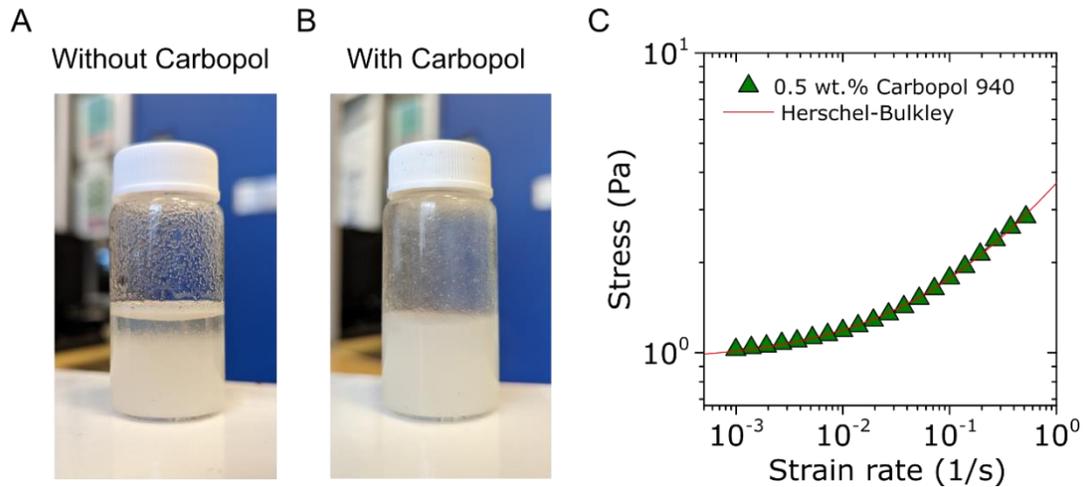

**Figure S2.** Yield stress fluids in the matrix prevent dextran particle sedimentation. Representative pictures of the composite hydrogels A) without Carbopol and B) with Carbopol (0.5 wt. %), in 20 mL scintillation vials. C) Flow curve of a 0.5 wt. % solution of Carbopol, fitted to the Herschel-Bulkley model.

**Dispersion of dextran particles in the composite hydrogels**

We find that dextran does not directly undergo particle-particle aggregation in our composite hydrogels. The large size of the dextran particles ($D \approx 120$ μm) and lack of specific interactions between two particles should result in a preferential interaction with the polymer matrix which offers multivalent binding sites for the particles – this is especially so for the cationic dextran particles. To demonstrate that the particles do not exhibit aggregation, we imaged $\phi_N = 0.01$ and $\phi_C = 0.01$ composite hydrogels which were gelled in a square capillary (Vitrocom) and sealed with a mixture of Vaseline-lanolin-paraffin. The images show that the particles are generally well-dispersed with minimal direct contact (Fig. S3). The particles do exhibit proximity however, which can be attributed to bridging interactions in the system. The images also show that particles exhibit minimal interpenetration with each other – a behavior which can result in non-linear elastic behavior from the particles even without the polymer matrix[3].

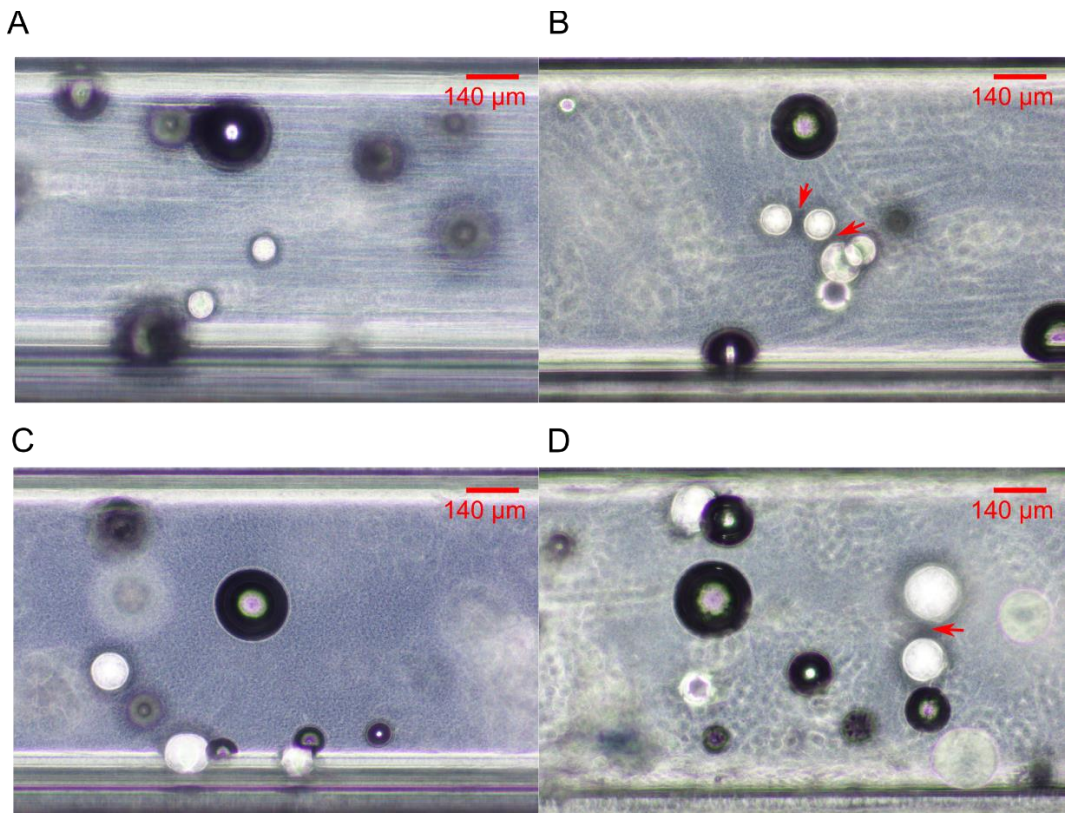

**Figure S3.** Representative microscope images (Olympus CKX53) of A,B) $\phi_N = 0.01$ and C,D) $\phi_C = 0.01$ composite hydrogels, all gelled in 0.9 mm width x 0.9 mm thickness square capillaries (Vitrocom). Note that the particles are in various states of focus due to the thickness of the capillary sample. Some of the particles in the same line of focus exhibit proximity without direct contact (red arrows), which may be attributed to polymer-induced bridging interactions.

## Compressibility of the composite hydrogels

We calculated the Poisson's ratio $v$ of our gel systems by tracking the transverse strain that arises during the quasi-static axial compression experiments. This was done through a camera setup placed under a clear rheometer plate, with gels dyed red using food coloring (Fig. S4A). The perimeters of the gels were then identified using MATLAB, and fitted to a circle to obtain the radius as a function of axial strain, and thus transverse strain. The Poisson's ratio $v$ was calculated via the relation $v = \lambda_{trans}/\lambda_{axial}$.

The composite hydrogels are almost completely compressible, with $v \sim 0$. The pectin hydrogels with Carbopol display a moderate amount of radial expansion ($v = 0.2$), but we find that this can be mostly attributed to the Carbopol, as pectin hydrogels without Carbopol also exhibits $v \sim 0$. The complete compressibility of the gels shown here is similar to that observed in biopolymer networks,[4-6] which would indicate that both the pectin and composite hydrogels exhibit fairly large pore sizes (despite the short persistence length of pectin, and despite the kPa scale elasticity of these systems).

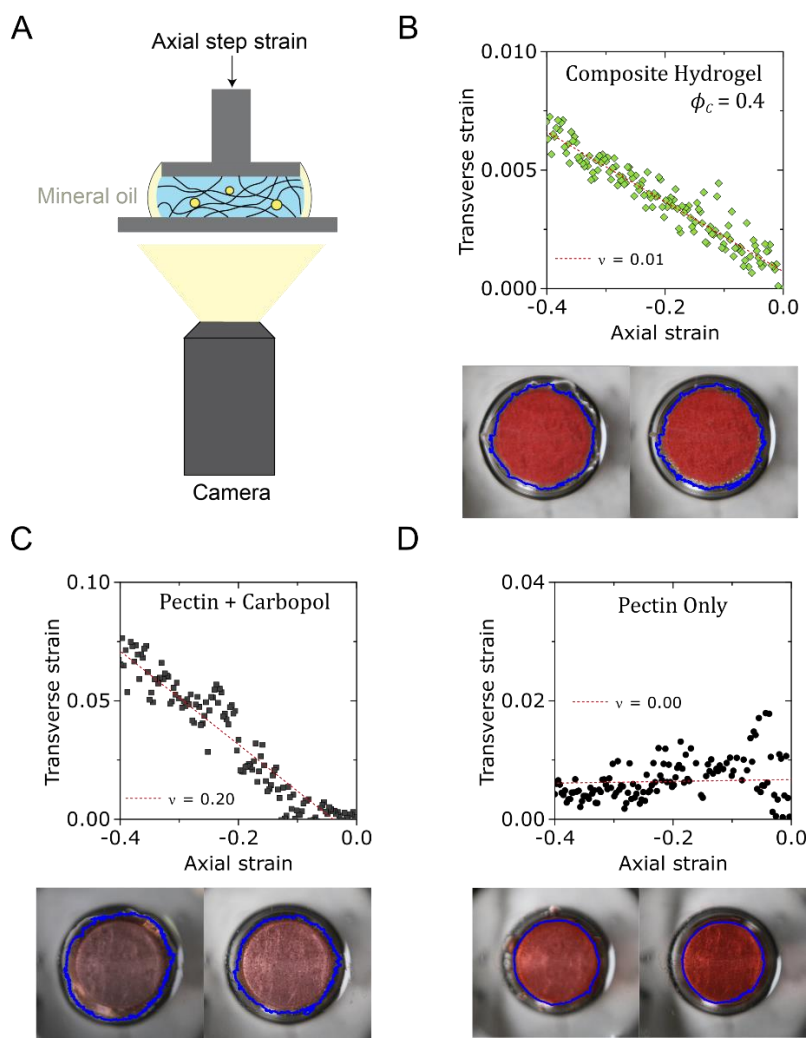

**Figure S4.** Analysis of gel compressibility under quasi-static axial strains. A) Schematic illustration of the setup used to analyze the radial expansion of the gel systems under compression. All experiments were done using a 20 mm parallel plate geometry (DHR-3). Transverse and axial strains of B) composite hydrogels with $\phi_C = 0.4$, C) pectin hydrogels with Carbopol, and D) pectin hydrogels without Carbopol. The slope was used to calculate the Poisson's ratio $v$. The images under each graph show the pictures of the gels – dyed red with food coloring – before (right) and after (left) the compression experiments. The blue outline illustrates the perimeter of the gel as identified using MATLAB, with which transverse strain is calculated.

**Stiffening of composite hydrogels under quasi-static compression**

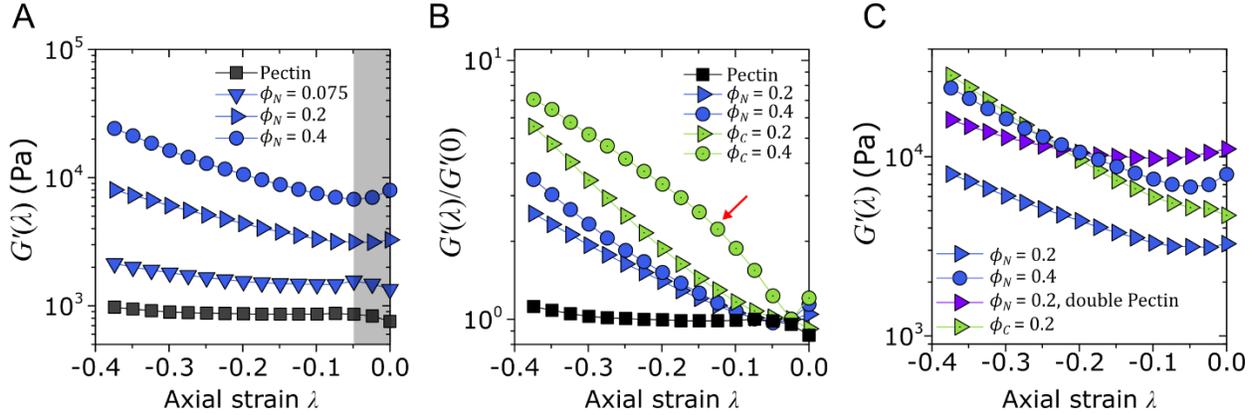

**Figure S5.** Representative storage modulus $G'$ as a function of compressive axial strain $\lambda$ for the composite hydrogel (all systems contain Carbopol). A) $G'(\lambda)$ of the pectin hydrogel, and of the composite hydrogel with $N$ particles at various volume fractions. Both the linear shear modulus, $G'(\lambda = 0)$, and the extent of compression stiffening are observed to increase with $\phi$. The anomalous behavior observed at low $\lambda$ are shaded in the panel. The initial stiffening behavior in the pectin and $\phi_N = 0.075$ may arise due to the relatively high Poisson ratio $\nu$ of the gels at $\phi_N$ (Fig. S4C), since this can cause buckling (which in turn can cause stiffening[7]) prior to sample expulsion. The initial softening behavior at $\phi_N = 0.2$ and $\phi_N = 0.4$ are expected behaviors arising from chain buckling in compressed biopolymer networks[4]. We normalize our data after the initial softening regime subsequent figures (panel B, Fig. 1C) to focus on the emergent stiffening behaviors in the composite hydrogels. B) The storage modulus values normalized by $G'(0)$ (the $G'$ value after the initial anomalous regime) for composite hydrogels with varying interaction strength $\varepsilon$. The hydrogels with charged particles exhibit more substantial compression stiffening across all $\phi$. At high $\phi$, there is a knee point in the slope of the curve (red arrow), which characterizes the onset of the jamming regime. C) Comparison of $G'(\lambda)$ values for composite hydrogels with varying $\varepsilon$ as well as polymer concentration. The $\phi_C = 0.2$ gel shows a greater compression stiffening effect compared to $\phi_N$ gels with higher $G'(\lambda = 0)$ (the $\phi_N = 0.4$ as well as $\phi_N = 0.2$ with 2 wt. % of pectin), showing that compression stiffening does not directly correlate with the elastic modulus of the gels.

**Compression stiffening master curves from repeat experiments**

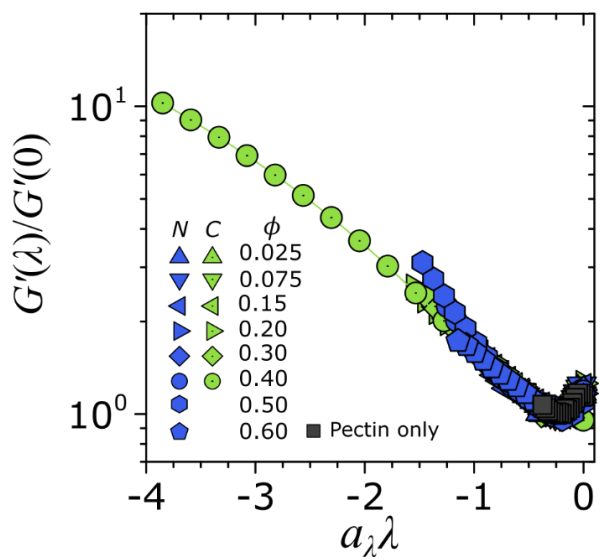

**Figure S6.** Second compression stiffening master curve from repeat experiments.

**Determination of the jamming threshold**

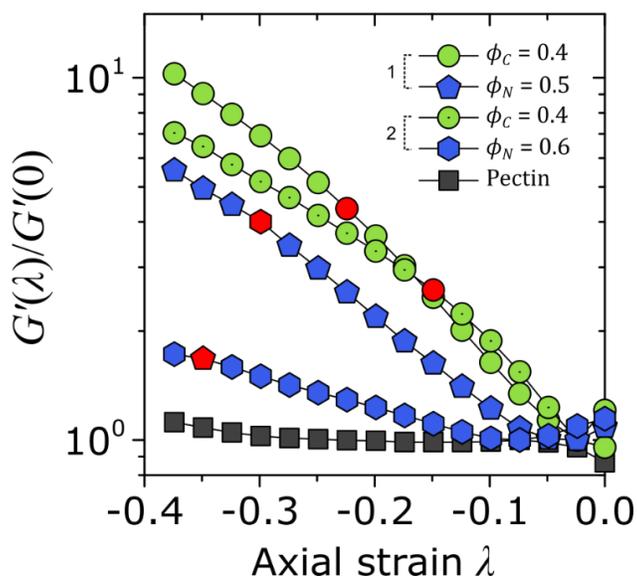

**Figure S7.** Normalized $G'$ of composite hydrogels at volume fractions $\phi$ exhibiting jamming behavior (across two sets of experiments per $\phi$). The jamming threshold strain $\lambda_{jam}$ is identified by calculating the knee point of each curve, and is highlighted in red for each curve.

**Fitting the linear elasticity of composite hydrogels with the Batchelor-Green hydrodynamic model**

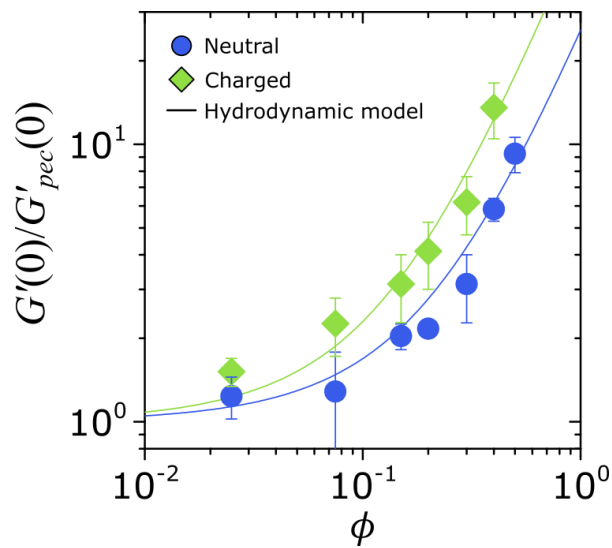

**Figure S8.** Same data as Fig. 1E, fitted instead to the Batchelor-Green ($c_2 = 5.2$) variant[8] of the hydrodynamic model in Eqn. 1. The resulting $\chi = \left(\phi_{jam}(N)/\phi_{jam}(C)\right)^{1/3} = 1.17$.

**Measurement of pectin adsorption capacity of neutral and charged particles**

To verify that the charged particles can indeed adsorb a greater concentration of pectin compared to the neutral particles (Fig. 1F), we dissolved 1 g of the $N$ or $C$ dextran particles in a 1 wt. % aqueous solution of pectin to facilitate polymer adsorption on the particles, filtered the particles using a grade 1 filter paper (Whatman, pore size < 11 $\mu m$), placed the filter paper with the particles in a $T = 100°C$ for 6 hrs, and weighed the samples after complete water evaporation. All results were normalized with those obtained from controls, which comprised of dextran particles dissolved in a pure aqueous solution without pectin, to obtain the adsorbed quantity of pectin on the particles. The results show a 42% greater adsorbed quantity of pectin with the $C$ particles compared to the $N$ particles, and support the molecular picture illustrated in Fig. 1F.

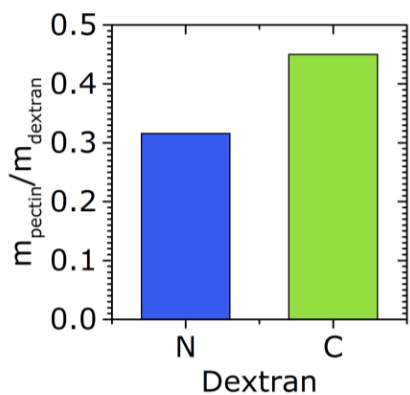

**Figure S9.** Adsorbed quantity of pectin on $C$ and $N$ particles after immersion in a 1 wt. % solution of pectin.

**Normal force characterization of the composite hydrogels during gelation**

The normal force response of the composite hydrogels is biphasic. The first phase involves an increase in the normal force, which can be attributed to the swelling-induced expansion of the dextran particles (Fig. S1). The second phase involves a decrease normal force, which is more commonly expected from the contraction that arises in gelling systems.[9] As our composite hydrogels form a percolated filler-polymer network, we can attribute the decrease in normal force to the extent of contraction experienced by the fillers during gelation. We indeed find that the charged particles exhibit a more significant drop in the normal force compared to the neutral particles, in support of the schematic shown in Fig. 1F in which stronger filler-polymer interactions mediate a bulk contraction of the gel, and thus an increase in the effective filler volume.

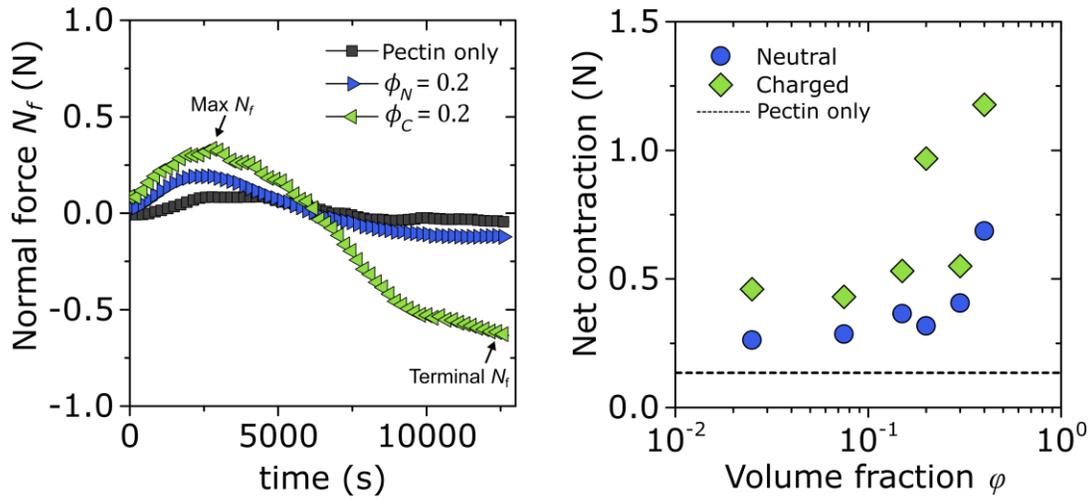

**Figure S10.** A) Representative evolution of the normal force of the composite hydrogels during gelation, and B) calculated net contraction of the composite hydrogels during gelation across various $\phi$. The net contraction is quantified by the difference between the maximum normal force and the terminal normal force (shown in panel A), i.e. Net contraction $=$ Max $N_f$ − Terminal $N_f$.

**Differential modulus characterization of pectin and composite hydrogels**

We characterized the non-linear elasticity of the pectin and composite hydrogels by first characterizing the differential storage modulus, in which the terminal slope of the elastic stress $\sigma'$ was measured for each Lissajous cycle (Fig. 2B,C), thus providing $G'_K(\gamma_0) = (d\sigma'/d\gamma)|_{\gamma_0}$. The analysis of this quantity for the pectin and composite hydrogels shows that the composite system exhibits a softer response than the matrix pectin hydrogel (Fig. S11A,B). We also followed the more conventional route to obtaining the differential storage modulus, where a small-amplitude oscillatory shear is superposed onto a step stress $\sigma_0$.[10-12] This form of the differential storage modulus, $K'(\sigma_0)$, was characterized by superimposing a $\sigma = 10\ Pa$ oscillatory shear at $\omega = 10$ rad/s onto the step stress $\sigma_0$, where the step stress was alternated from an increasing step stress to a zero step stress at 20 s intervals. In general, we found this method to be the least suitable for characterizing the non-linear elasticity of our systems, as failure (or slip) occurs much earlier than from LAOS-based methods (Fig. S11C). We hypothesize that this may be due to the greater overall work put into the system by the step stress protocol compared to LAOS, and due to the fact that our materials are much stiffer than the biopolymer networks which are typically studied using this method.

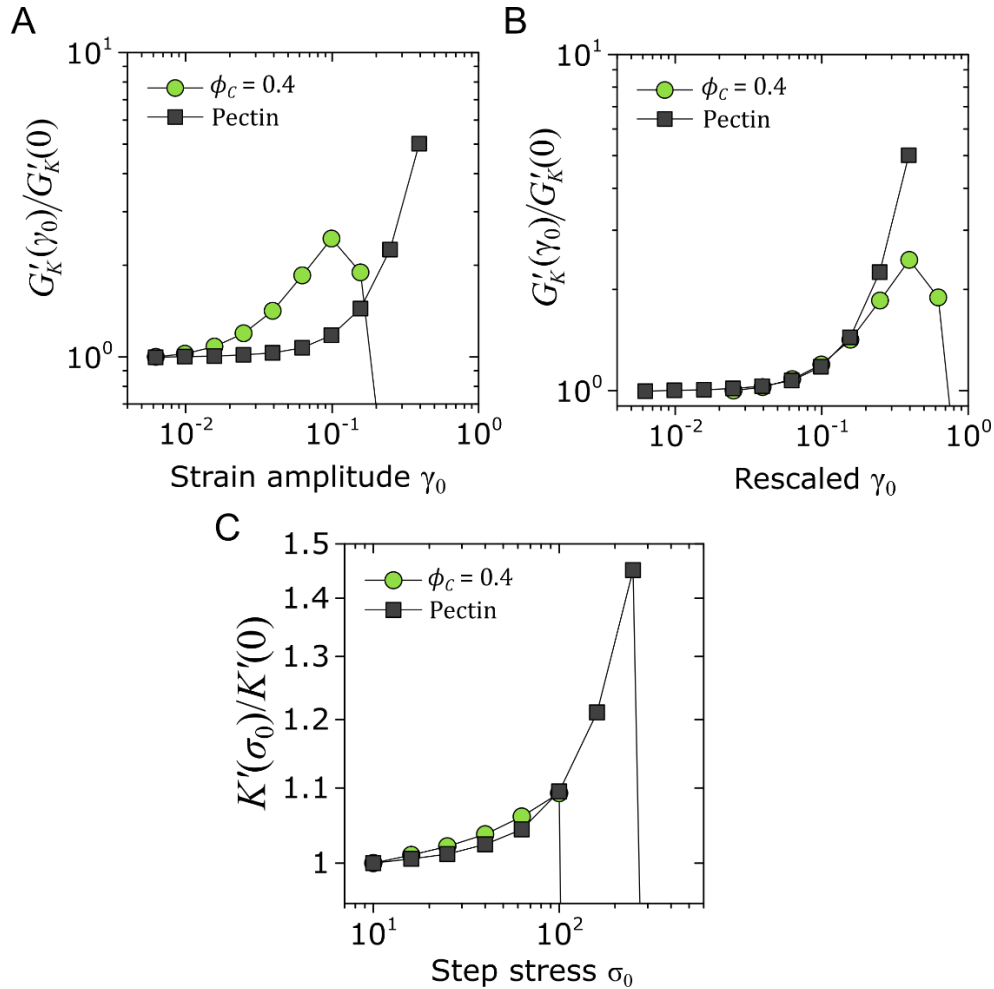

**Figure S11.** A) Normalized differential storage modulus $G'_K(\gamma_0)$ of the pectin and $\phi_C = 0.4$ composite hydrogels. B) Same data, but the $\phi_C = 0.4$ data is horizontally shifted onto the pectin data. C) Normalized differential storage modulus $K'(\sigma_0)$ obtained using the step stress protocol.

**Filler-polymer interaction-dependent Mullins effect in the composite hydrogels**

We performed explicit characterization of the Mullins effect, which refers to the stress softening that is observed in composite systems in response to increasing cyclic strain (Fig. S12A). A distinguishing feature of systems exhibiting the Mullins effect is that the loading curves of the subsequent cycle traces the unloading curve of the previous cycle, resulting in measurable quantities such as dissipation (purple arrows, Fig. S12B,C) and plastic deformation (purple arrows, Fig. S12D,E).[13-15] The high-$\varepsilon$ composite hydrogels with $C$ fillers exhibit significantly greater Mullins effect than those with $N$ fillers, indicating that strong filler-polymer interactions result in greater polymer adsorption on the fillers, and thus greater elastoplastic dissipation.[16]

As the LAOS experiments also involve increasing cycling strain, we should be able to observe some traits of the Mullins effect in LAOS for the $C$-filled hydrogels. We indeed find that the maximum shear strain of the previous cycles approximately corresponds to the knee point of the subsequent elastic stress curves (Fig. S12F) (they are in fact slightly higher, which is consistent with Fig. S12B). We also measured transient Lissajous curves (i.e., as opposed to the steady state curves presented in the paper) and indeed find that loading curves of subsequent cycles trace the unloading curves of the previous cycles (Fig. S12G).

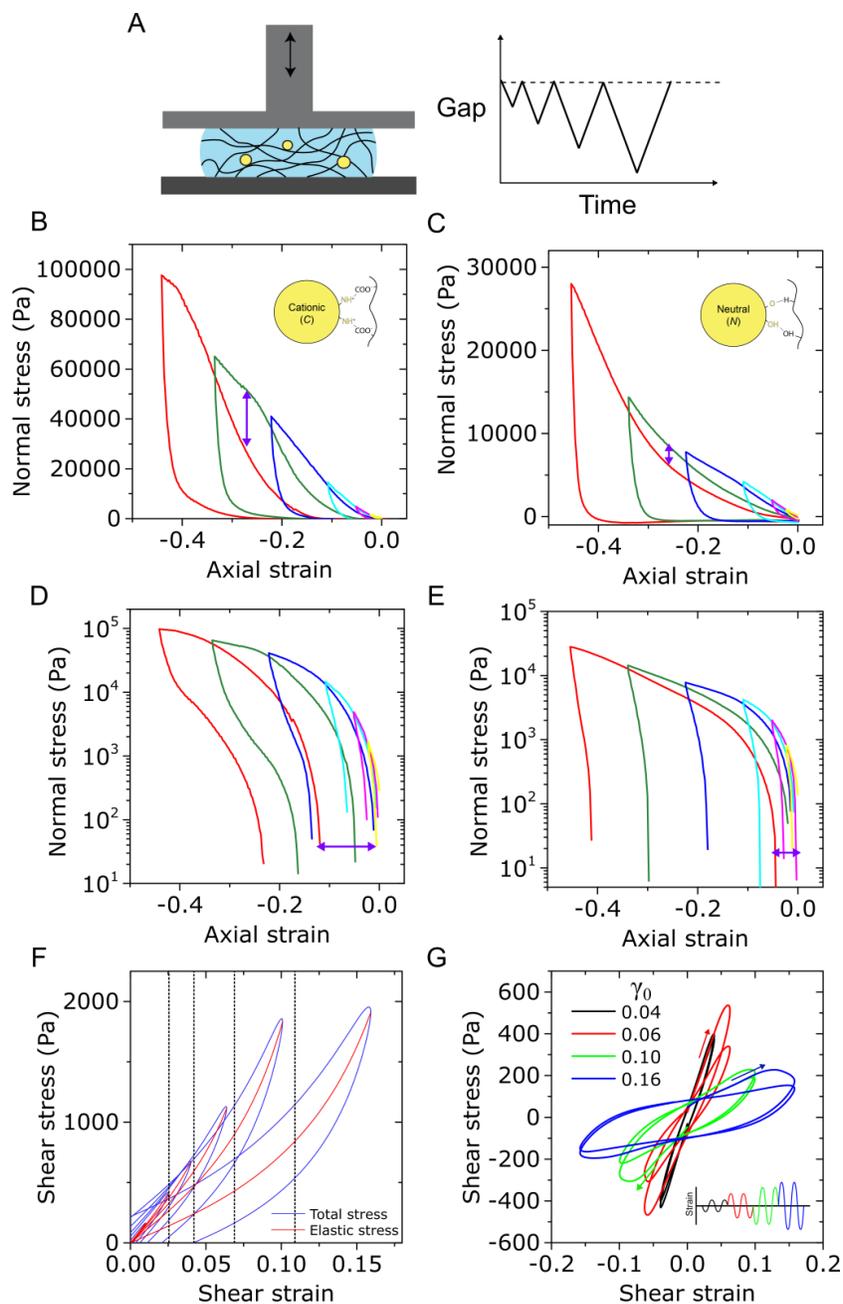

**Figure S12.** Mullins effect characterization of the composite hydrogels. A) Method of analysis. Composite hydrogels with $N$ and $C$ fillers were gelled in disc molds (25 mm diameter, 3.6 mm thick), after which cyclic compression experiments were conducted on a DHR-3 rheometer (40 mm parallel plate) at a speed of 0.01 mm/s. B,C) Compression stress-strain curves of the $C$ and $N$ filled systems, and D,E) the same plots in semilog axes. The Mullins effect results in dissipation (vertical purple arrows in B,C) and permanent strain (horizontal purple arrows in D,E). F) Lissajous curves of the $\phi_C = 0.4$ gel measured at $\gamma_0 = 0.04, 0.064, 0.10$, and $0.16$. The dashed lines show the knee points of the elastic stress curves. G) Transient Lissajous curves of the $\phi_C = 0.4$ gel measured at $\gamma_0 = 0.04, 0.064, 0.10$, and $0.16$. Each curve shows two cycles, after which the strain direction is flipped (the method is illustrated in the bottom right inset). The arrows show the loading direction of the curves, roughly tracing the unloading curves of their previous curves.

**Effects of sandpaper on compression and shear rheological characterization**

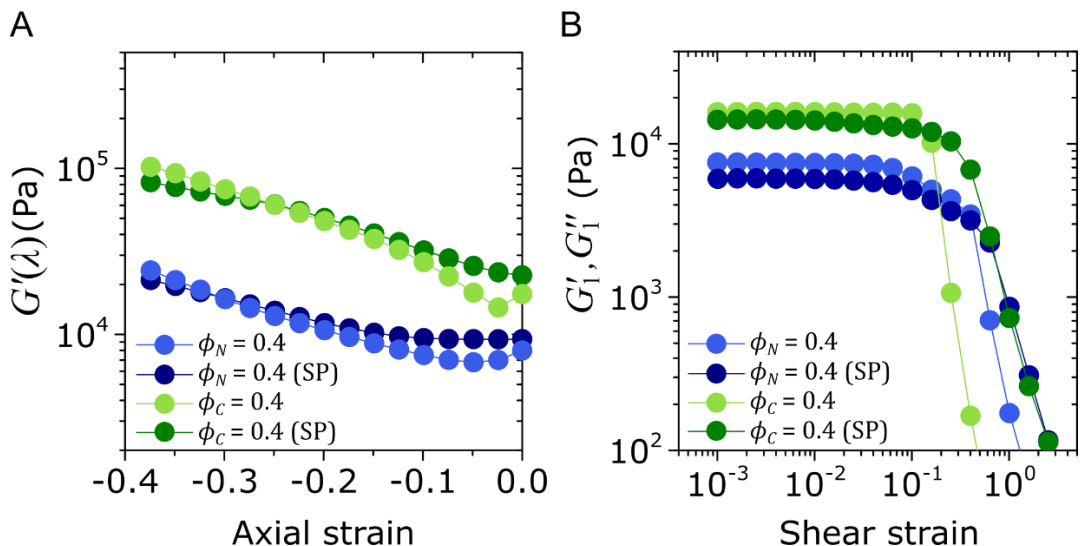

**Figure S13.** Representative A) compression and B) shear rheological characterization for the composite hydrogels with 60 grit sandpaper (SP) on the rheometer plates. Most experiments in the manuscript are done without sandpaper as we found the difference in the results to be relatively minimal.

**Characterization of the occurrence of slip during LAOS experiments of composite hydogels**

Understanding the role of slip is critical for ensuring that slip effects are not influencing the shear responses of the composite hydrogels, such as the horizontal translation in $\sigma'(t)$ and the shape of the $G'_K(\gamma)$ curves. We find that the $N$-filled composite hydrogels at high $\phi$ exhibit clear signatures of slip, as there is a gradual drop in the $G'$ preceding the yielding point (Fig. S14A).[17] We also find that the corresponding $G'_K(\gamma)$ curves in the slip regime for $\phi_N = 0.4$ begin to deviate from the $G'_K$ responses of the polymer matrix as well as the $\phi_C = 0.4$ system (Fig. S14B).

We thus find a curious result where the higher $\varepsilon$ composite hydrogels are both cohesive (higher elasticity) *and* adhesive (higher slip resistance) – two properties which are conventionally thought to be diametric in adhesion literature. We hypothesize that this can occur because the $C$ fillers are cationically charged, are present on the gel surface, and thus provide greater adhesion to the metal plates of the rheometer (Fig. S14C). To validate this idea, we placed a 10 µL drop of a mineral oil solution with 2 wt. % of the anionic surfactant AOT, and we indeed find that the droplet completely wetted the $C$-filled hydrogel, while remaining beaded on the $N$-filled hydrogel (Fig. S14D). Moreover, to validate that these surface interactions affect the slip response of the gels, we repeated the LAOS experiments by placing poly(vinyl chloride) (PVC) tape on the rheometer plates, since these plastic surfaces should not facilitate charge-induced adhesion. We find that the $N$-filled hydrogels exhibit the same LAOS response, and the $C$-filled hydrogels now exhibit a dramatic slip response characterized by a lowering in the $G'$ (Fig. S14E,F). The analysis of the $G'_K$ response in this slip regime shows that $G'_K$ again deviates from the expected response of the system (Fig. S14G).

Overall, we find that slip results a *lowering* of the $G'_K(\gamma)$ response. This provides a useful context to interpret the tissue experiment results, as while we find that $G'$ and $G''$ drop as a function of $\gamma_0$ (Fig. 3C), the calculated $G'_K$ exhibit a stress scaling of $G'_K(\sigma) \sim \sigma^{1.5}$ which is the expected upper limit of the scaling of $G'_K$ arising from the biopolymer matrix. This suggests that LAOS characterization of tissues are likely not convolved by slip.

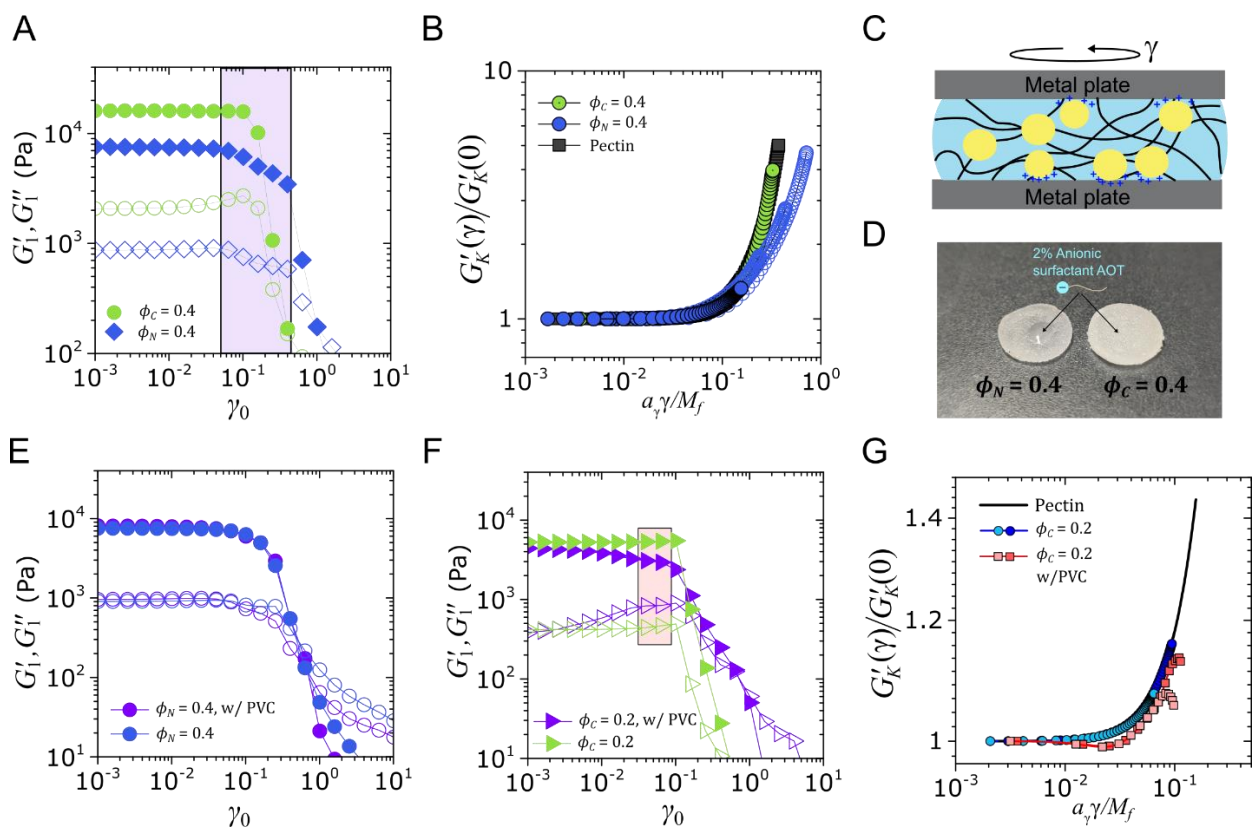

**Figure S14.** Characterization of slip effects on the composite hydrogels ($\phi = 0.4$). A) First harmonic storage and loss moduli of the composite hydrogels (same data as Fig. 2A), highlighting the region exhibiting slip in the $\phi_N = 0.4$ system. B) Shear stiffening master curve of the pectin and composite hydrogels, where the data for each system is shifted by $1/M_f$ and then all of the data is shifted onto the pectin curve. The hollow symbols are $G'_K(\gamma)$ of $\phi_N = 0.4$ gels extracted from the slip region in panel A. C) Schematic illustration of the charge-induced adhesion of the gels onto the rheometer plates facilitated by $C$ fillers. D) Pictures of composite hydrogel discs (25 mm diameter) with a droplet of a solution of mineral oil with 2 wt. % AOT. E,F) First harmonic storage and loss moduli of the composite hydrogels with PVC tape on the rheometer plates. G) $G'_K(\gamma)$ of the $C$-filled composite hydrogels in the slip region highlighted in panel F using PVC tape, shifted onto the pectin data.

**Collapse of shear stiffening curves for pectin and composite hydrogels**

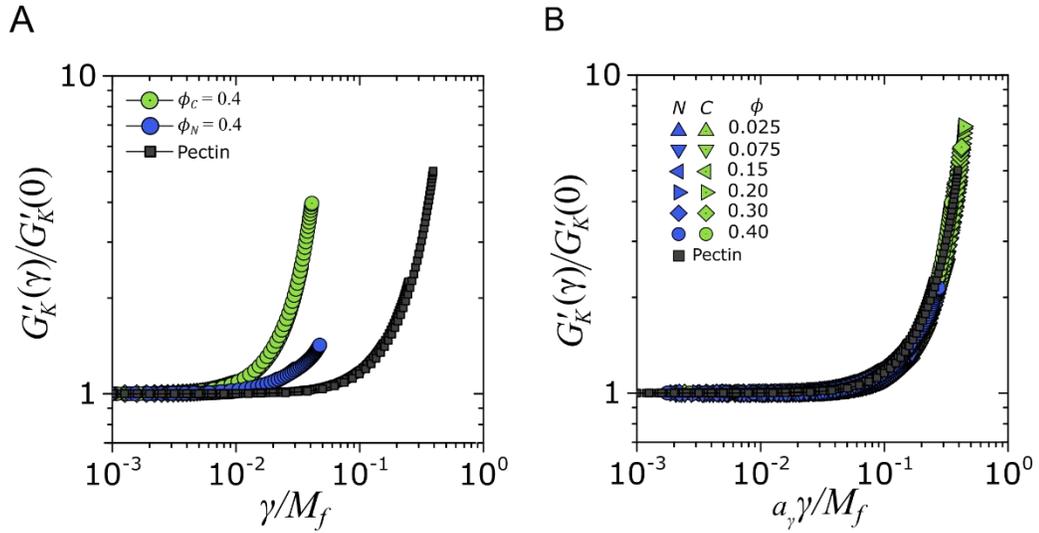

**Figure S15.** A) Representative examples of the $G'_K(\gamma)$ curves after performing horizontal shifting to account for the Mullins effect (see Fig. 2C,D). The curves are self-similar, but horizontally translated. B) Collapse of $G'_K(\gamma)$ as a function of $\phi$ and $\varepsilon$ using the shear strain amplification factor $a_\gamma$.

**Shear stiffening master curves from repeat experiments**

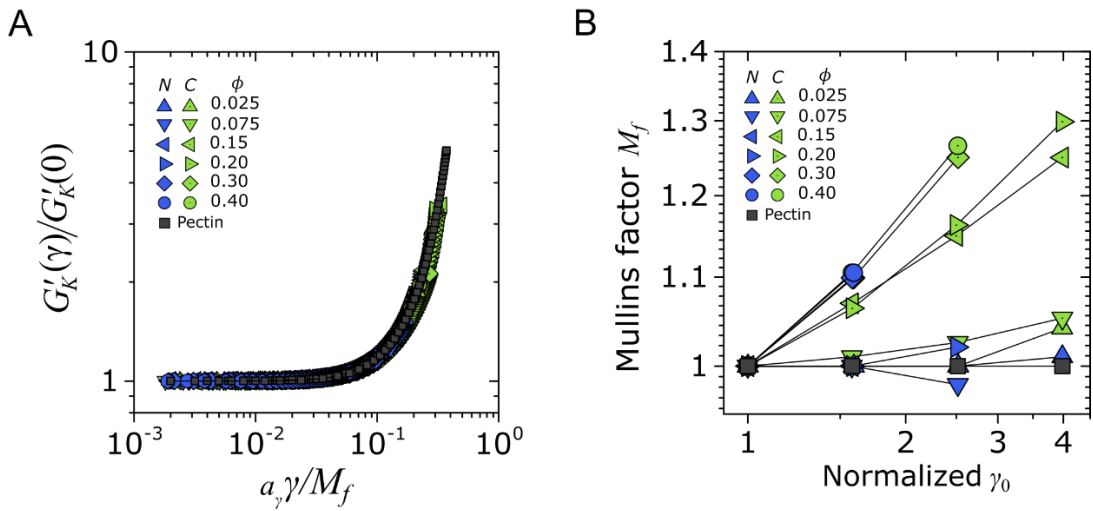

**Figure S16.** Results from repeated shear experiments. A) Master curve of $G'_K(\gamma)$ as a function of $\phi$ and $\varepsilon$. The associated shear strain amplification factor $a_\gamma$ is used to obtain statistics for Fig. 2H in the main manuscript. B) The Mullins factor $M_f$ for the hydrogels as a function of $\phi$ and $\varepsilon$.

**Comparison between strain amplification and elasticity**

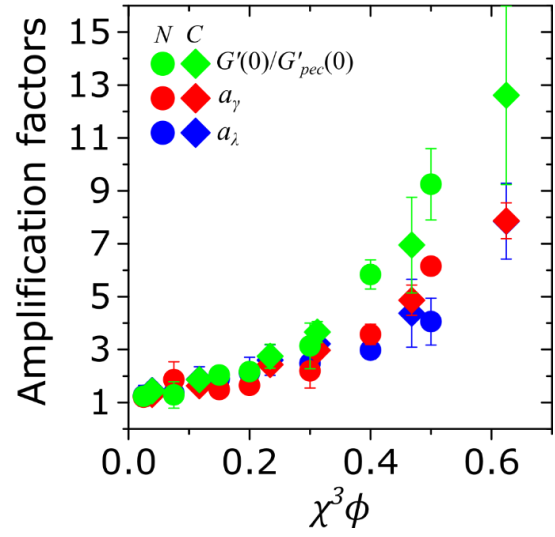

**Figure S17.** Comparison between the compression and shear strain amplification factors $a_\lambda$ and $a_\gamma$, and the linear elasticity of the system $G'(0)/G'_{pec}(0)$ (shown in linear scale). The strain amplification factors provide a reasonable description of $G'(0)/G'_{pec}(0)$ at $\chi^3\phi < 0.2$, but begin to deviate at higher $\chi^3\phi$.

**Representative pictures of tissue samples studied**

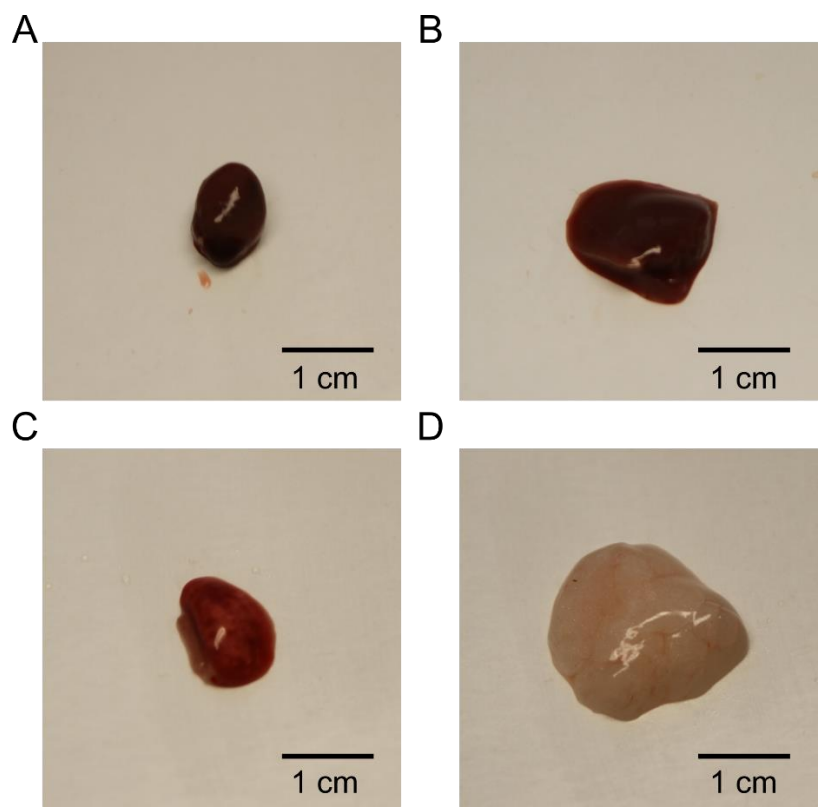

**Figure S18.** Representative pictures of A) heart, B) liver, C) lung, and D) adipose samples excised from mice.

**Compressibility of tissues**

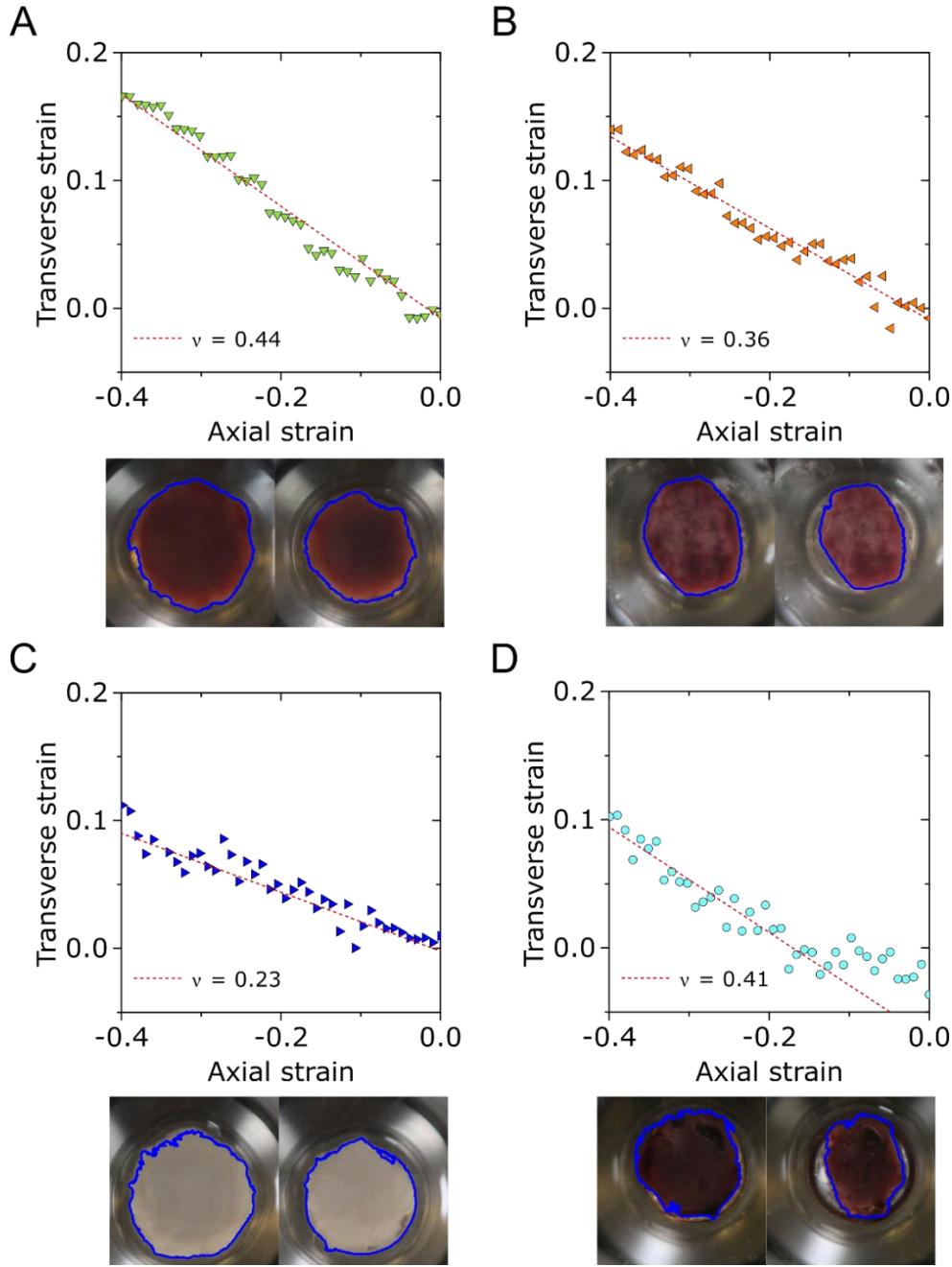

**Figure S19.** Analysis of tissue compressibility (following protocols illustrated in Fig. S4) for A) liver, B) lung, C) adipose, and D) heart. The Poisson's ratio is calculated by the relation $v = \lambda_{trans}/\lambda_{axial}$. The heart sample exhibits a non-linear dependence of transverse strain on axial strain. We find this arising due to the flattening of the heart sample at low strains (the sample is initially round (Fig. S18)); the Poisson's ratio is thus calculated after the initial flattening stage.

**Elastic stress dependence of the differential modulus of pectin and composite hydrogels**

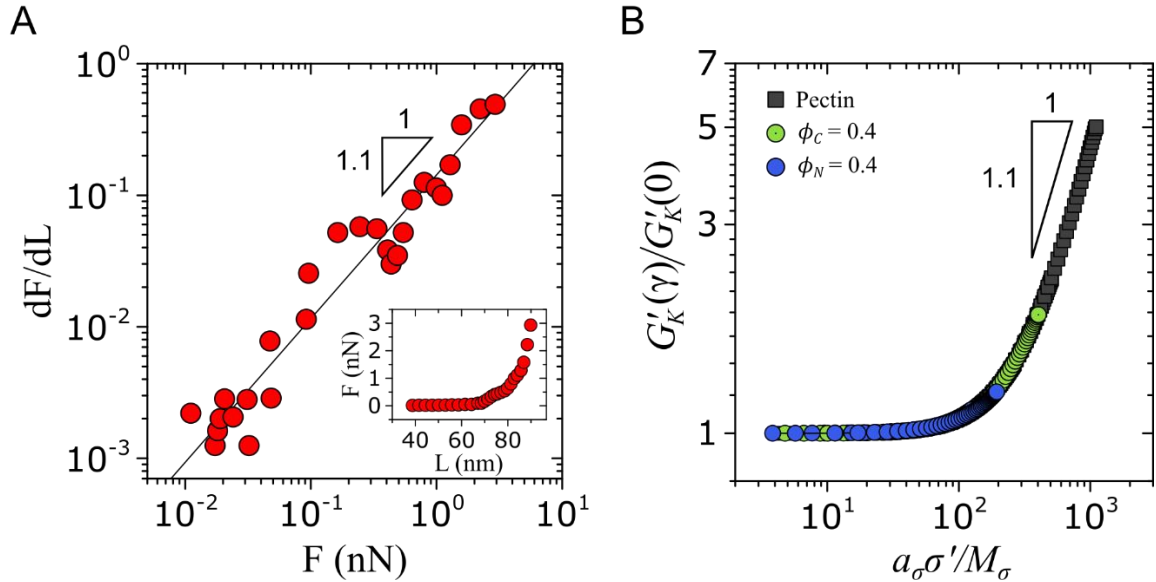

**Figure S20.** The non-linear elasticity of pectin and composite hydrogels can be attributed to the stretching of the pectin chains. A) Derivative of force F (as a function of displacement L) versus F of pectin hydrogels. Fitting the data to the function $dF/dL \sim F^z$ yields $z = 1.1$. The inset shows the raw F-L data of pectin, retrieved from reference[18]. B) The strain-dependent differential modulus of pectin and composite hydrogels plotted as a function of $\sigma'$. The terminal slope of the curve follows $G'_K(\gamma) \sim \sigma'^z$. Note that the horizontal axis is rescaled by the stress-induced Mullins effect and the stress amplification factor (see Fig. S21 for discussion about the curve shifting process).

## Shifting protocol to obtain elastic stress dependent master curves

The strain-dependent differential storage modulus $G'_K(\gamma)$ can also be explored as a function of elastic stress $\sigma'$ in similar vein to the classical literature on this topic.[10,19] The $G'_K(\gamma)$ vs $\sigma'$ curves exhibit horizontal translation with increasing shear strain amplitude $\gamma_0$ – in similar vein to $G'_K(\gamma)$ vs $\gamma$ curves (Fig. 2D,E) – but the trend is reversed where the curves now show an earlier onset of stiffening with elastic stress (Fig. S21A). This indicates that a lower amount of stress is required to strain the polymer network to its non-linear regime due to accumulated damage. The associated Mullins shifted factor, $M_\sigma$, appears to be inversely related to $M_f$ (Fig. S21B). The resulting master curves for the different tissues (Fig. S21C) can then be rescaled into a master curve (Fig. 3E), through a stress amplification factor $a_\sigma$. This stress amplification factor qualitatively appears to be inversely related to the elasticity of the system (Fig. S21D).

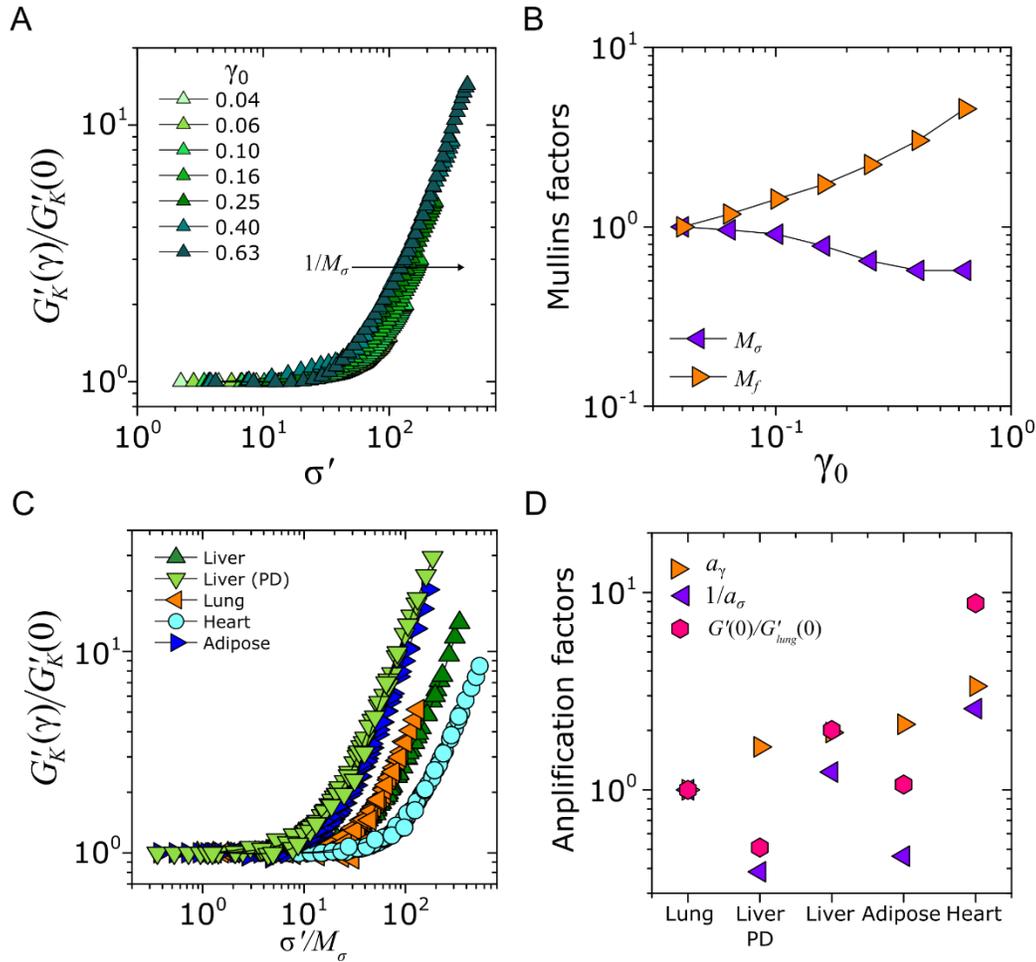

**Figure S21.** A) Representative $G'_K(\gamma)$ data for heart as a function of $\sigma'$. The data is shifted to the reference curve ($\gamma_0 = 0.04$) through the shift factor $1/M_\sigma$. B) Comparison of the strain-based Mullins factor $M_f$ and the stress-based Mullins factor $M_\sigma$ for the heart sample, showing an approximate inverse (but not linearly inverse) relationship with each other as a function of strain amplitude $\gamma_0$. C) The $G'_K(\gamma)$ vs $\sigma'$ master curves for the different tissue samples. The shown curves are then horizontally shifted to the lung master curves with the shift factor $a_\sigma$. D) Comparison of the strain amplification factor $a_\gamma$, stress amplification factor $a_\sigma$, and elastic modulus $G'(0)$ normalized by that of the lung system.

**Shear strain stiffening of tissues**

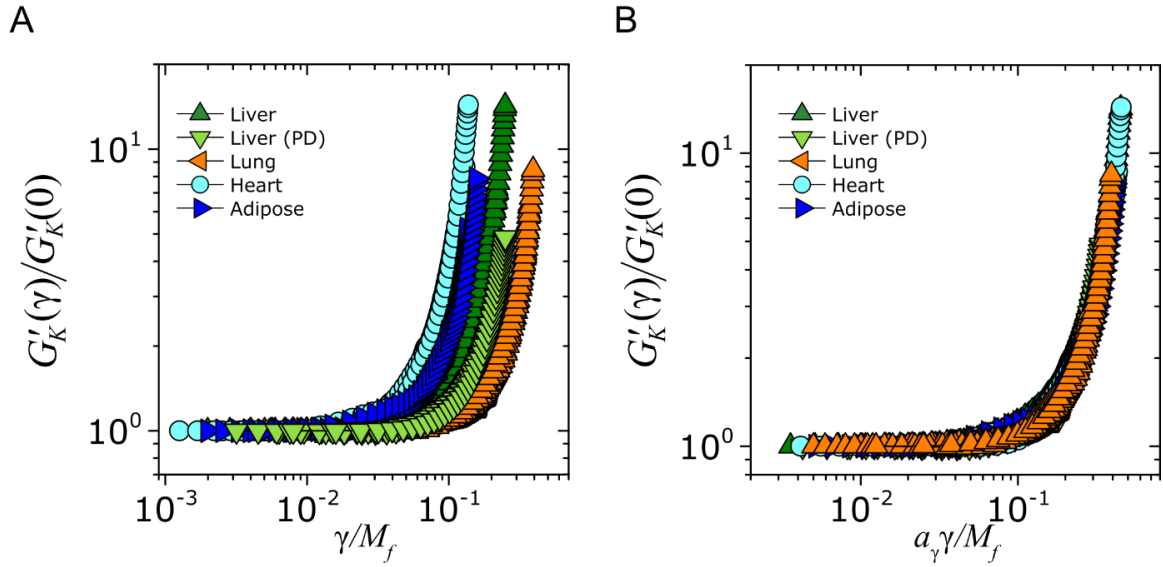

**Figure S22.** Strain-dependent differential storage modulus $G'_K(\gamma)$ as a function of $\gamma$, A) before and B) after shifting with the strain amplification factor $a_\gamma$.

**Results from repeat experiments on tissues**

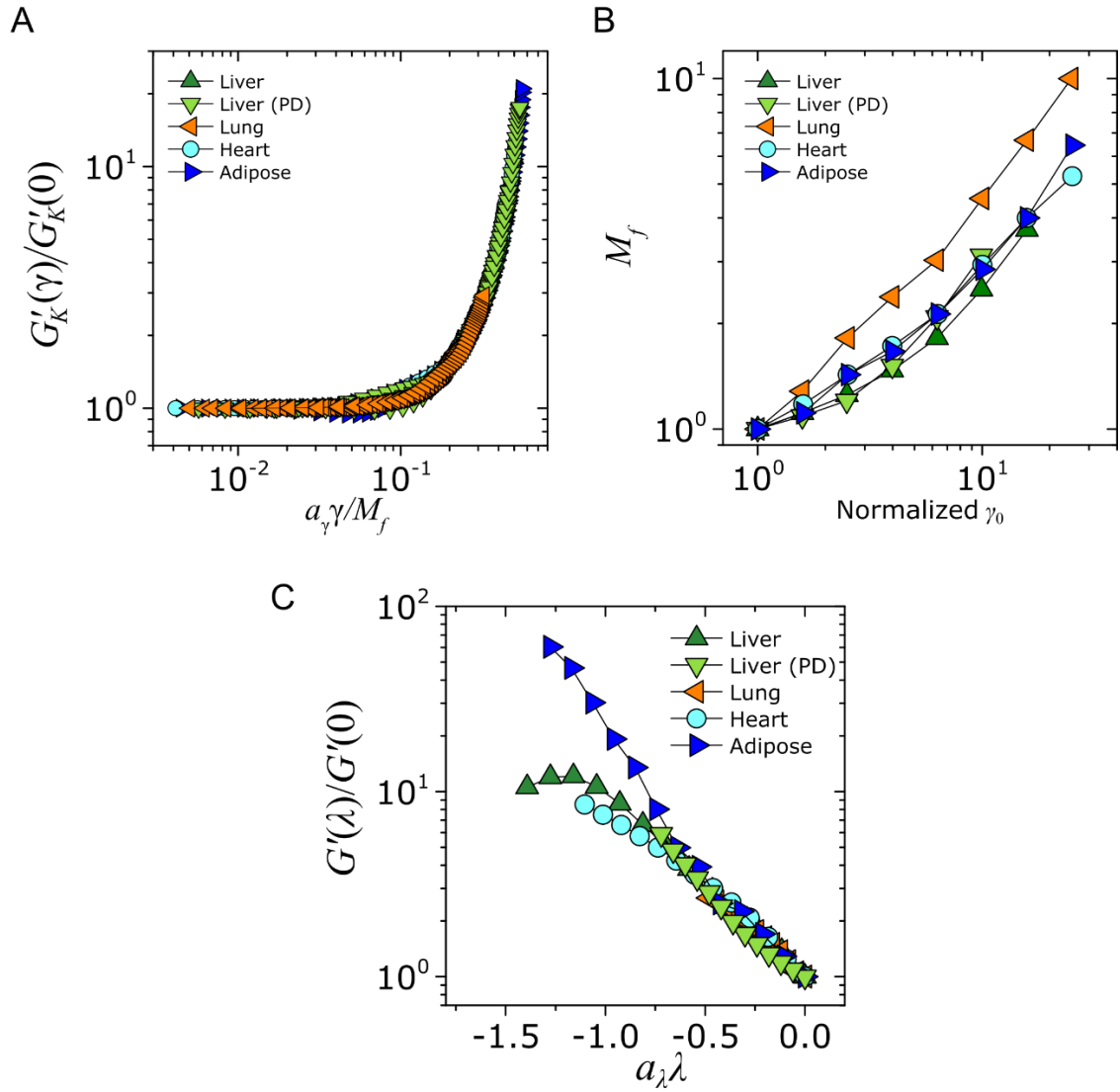

**Figure S23.** Results from repeats of shear and compression stiffening tests on tissues. A) Shear stiffening master curve, and B) associated Mullins factor $M_f$ from the repeat experiments. C) Compression stiffening master curves. The data shown here are used to obtain statistics for Fig. 3H.

**Representative gelation plot of pectin and composite hydrogels**

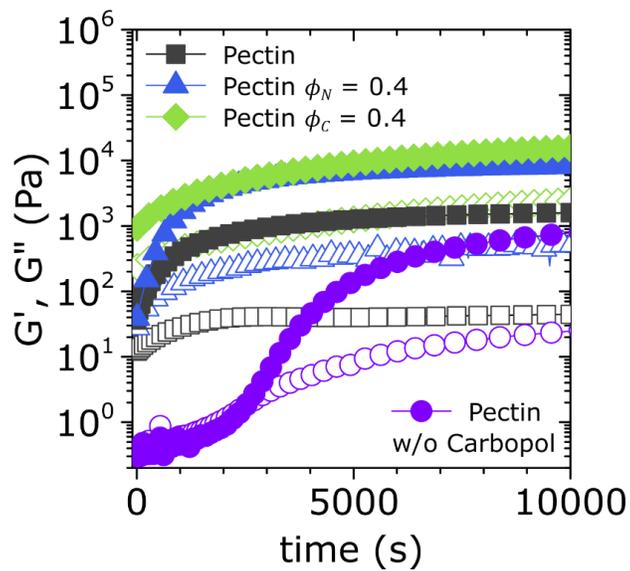

**Figure S24.** Gelation plot of the pectin hydrogels (with and without Carbopol additives), as well as composite hydrogels with varying $\varepsilon$.